\documentclass{JHEP3}

\usepackage{epsfig}
 
\newcommand \Pomeron {I\!\!P}
\newcommand{\lsim}{\mbox{\,\raisebox{.3ex}
    {$<$}$\!\!\!\!\!$\raisebox{-.9ex}{$\sim$}\,}}
\newcommand{\gsim}{\mbox{\,\raisebox{.3ex}
    {$>$}$\!\!\!\!\!$\raisebox{-.9ex}{$\sim$}}\,}

\title{Nuclear shadowing in deep inelastic scattering on nuclei:
leading twist versus eikonal approaches}
\author{L.~Frankfurt\\
        Nuclear Physics Dept., School of Physics and Astronomy,\\
        Tel Aviv University, 69978 Tel Aviv, Israel\\
        E-mail: \email{frankfur@lev.tau.ac.il}}
\author{V.~Guzey\\
Special Research Centre for the Subatomic Structure of
Matter (CSSM),\\
Adelaide University,  5005, Australia,\\
\email{vguzey@physics.adelaide.edu.au}}
\author{M.~McDermott\\
        Division of Theoretical Physics,\\
        Dept. Mathematical Sciences, Liverpool University\\
        Liverpool L69 3BX, England\\
        E-mail: \email{martinmc@amtp.liv.ac.uk}}
\author{M.~Strikman\\
        Department of Physics, Penn State University,\\
        University Park, PA, 16802-6300, USA.\\
        E-mail: \email{strikman@phys.psu.edu}}

\abstract{
We use several diverse parameterizations of diffractive parton distributions, 
extracted in leading twist QCD analyses of the HERA diffractive deep inelastic scattering (DIS) data, to make predictions for leading twist nuclear shadowing of nuclear quark and gluon distributions in DIS on nuclei.
We find that the HERA diffractive data are sufficiently precise to allow us to predict
large nuclear shadowing for gluons and quarks, unambiguously.
We performed detailed studies of nuclear shadowing for up and charm sea quarks and gluons within several scenarios of shadowing  and diffractive slopes, as well as at central impact parameters.
We compare these leading twist results with those obtained from 
the eikonal approach to nuclear shadowing (which is based on a very different 
space-time picture) and observe sharply contrasting predictions 
for the size and $Q^2$-dependence of nuclear shadowing. 
The most striking differences arise for the interaction of small dipoles 
with nuclei, in particular for the longitudinal structure function $F_{L}^{A}$.}

\preprint{LTH-522, ADP-01-41/T473}

\begin{document}

\section{Introduction}

The high-energy scattering of leptons and hadrons from nuclear targets offers a unique
possibility to study the distribution of quarks and gluons in nuclei, i.e.,
large parton densities in a large volume.
It is now firmly established by several experiments
(EMC and NMC at CERN, a number of experiments at SLAC, E665 at Fermilab)
that at small values of Bjorken $x$, $5 \times 10^{-3} \leq x \leq 0.03-0.07$ (for $Q^2 \ge 1$ GeV$^{2}$), the
inclusive nuclear structure
 function, $F_{2}^{A}$, for DIS on a nucleus with $A$ nucleons, is
smaller than the incoherent sum of the nucleon
structure functions, $A F_{2}^{N}$. This phenomenon
is called nuclear shadowing (for recent reviews see \cite{Piller}).
It is important to emphasize that
the available data, which admittedly has a rather limited kinematic range,
$Q^2 \lsim 4$~GeV$^2$, demonstrate no significant $Q^2$-dependence of nuclear
shadowing. In addition, nuclear shadowing of the antiquark distributions has been observed
in the Drell-Yan process \cite{DY}, in the range $0.02~\lsim ~x ~\lsim ~0.05$
and $16 ~\lsim ~Q^2 ~\lsim ~35$~GeV$^2$. In other words, the data are consistent with the 
dominance of leading twist shadowing.

Why is it still interesting to work on the subject of nuclear shadowing in
DIS on nuclei after the topic has been investigated so thoroughly? There are 
several reasons.

Firstly, an accurate evaluation of nuclear shadowing is vitally important for both
light and heavy nuclei to pin down the nuclear parton 
distributions.
These form the boundary conditions for hard processes in electron-nucleus DIS 
experiments proposed in the USA (the electron-ion collider (EIC)) \cite{EIC} and
at DESY \cite{HERAnuc}, in nucleus-nucleus and nucleon-nucleus
collisions presently under way at BNL, and in proton-nucleus and nucleus-nucleus
collisions planned for the LHC at CERN.

Secondly, it is important to understand the intricate interplay of perturbative 
and non-perturbative effects in nuclear shadowing and its role in the parton 
distributions of nuclei at small $x$. Understanding this interplay is a key element 
in determining when the leading twist approximation (implicit in the DGLAP
evolution equations) breaks down for sufficiently small $x$.

Thirdly, because of the deep connection between nuclear shadowing
and diffraction in DIS on a nucleon established by
V.~Gribov \cite{Gribov}, accurate measurements of nuclear
shadowing should help to discriminate between competing models of diffraction.

The dynamics of nuclear shadowing in DIS at high energies is most transparent in the target rest frame where the virtual photon-nucleus (nucleon) scattering,
$\gamma^* (q) + A (P_{A}) \to X$, is a time-ordered three-stage process.

Firstly, the photon may be considered to fluctuate into a linear superposition
of hadronic components labelled by the basis set $|h_{k}\rangle$.
The components consist of quarks, antiquarks and gluons at small
relative transverse distances (and their associated color fields)
and of hadronic bound states at large transverse distances.
For large center-of-mass energy, $W$ ($W^2 = (q + P_A)^2$), i.e., small Bjorken $x \approx AQ^2/W^2$, the fluctuation occurs a long distance upstream of the nuclear target.
The incoming virtual photon (more precisely, the hadronic part of the virtual photon) state $|\gamma^{\ast} \rangle$ may be expressed
generically as follows
\begin{equation}
|\gamma^{\ast} \rangle = \sum_{k} c_{k} |h_{k}\rangle \ ,
\label{dec1}
\end{equation}
in which the index ``{\it k}'' should be understood to be a label which
fully specifies the hadronic state (i.e.,
it includes information about momentum fractions and transverse momentum
of the constituent partons, their helicities, etc.) and $c_{k} = \langle h_k | \gamma^* \rangle$.

Secondly, the hadronic configurations $|h_{k}\rangle$ interact strongly with
the nucleus
\begin{equation}
\sigma_{\gamma^{\ast} A}=\sum_{k,k^{\prime}} c^{*}_{k'}
 {\hat \sigma}_{k, k^{\prime}}^{A} c_{k} \, ,
\label{dec2}
\end{equation}
where the subprocess cross section, ${\hat \sigma}_{k, k^{\prime}}^{A}$, is related
via the optical theorem to the imaginary part of the forward scattering amplitude
for the process $h_{k}+A \to h_{k^{\prime}}+A$ normalized such that
${\hat \sigma}_{k, k}^{A}=\sigma_{h_{k}A}$, the total $h_{k}$-nucleus cross section.

Finally, the hadronic final state $X$ is formed.

In general, each state $h_{k}$ suffers attenuation: at sufficiently high energies
the states $|h_{k} \rangle$ interact coherently with several nucleons of the target.
These multiple interactions decrease ${\hat \sigma}_{k, k^{\prime}}^{A}$ relative
to the sum of ${\hat \sigma}_{k, k^{\prime}}^{N}$ of the individual nucleons,
which can be translated to the corresponding inclusive structure functions as
$F_{2}^{A}/(AF_{2}^{N})<1$. Thus, the coherent interaction of the photon with several
nucleons of the target nucleus invariably results in nuclear shadowing.

Equations~(\ref{dec1},\ref{dec2}) are very general. Any particular realization requires some
modelling of the hadronic content of the (virtual) photon and of the interaction with the target.
There are several models in the literature including Generalized Vector Meson Dominance (GVMD) models \cite{GVMD}, two-phase models including vector mesons and the Pomeron \cite{MT} or other scaling contributions \cite{Piller2}, and the QCD-improved aligned jet model \cite{AJM}.
In this paper, we review and compare the two most recent approaches to nuclear shadowing: the leading twist and eikonal approaches.

The leading twist approach was developed in a
recent work by Frankfurt and Strikman \cite{FS99}, who attempted
to place the treatment of nuclear shadowing on firmer ground. The authors
generalized the Gribov theory of nuclear shadowing in hadron-nucleus
scattering \cite{Gribov} to individual quark and gluon nuclear
parton densities in light nuclei, probed in DIS on nuclei.
As in the Glauber model, nuclei were treated as a dilute nucleon gas.
The connection between nuclear shadowing in inclusive DIS on the nucleus and 
diffraction in DIS on the proton, the factorization theorem for hard diffraction 
\cite{Collins} and high accuracy DIS diffractive data allowed the leading twist 
(LT) component of the nuclear shadowing correction to nuclear
parton distributions to be evaluated reasonably well. 
One immediate and very distinct consequence of the theory is that, at some initial scale $Q_{0}$ ($Q_{0} =2$~GeV in \cite{FS99}), nuclear shadowing for gluons is much larger than 
that for quarks (by a factor of about two or three for the lightest nuclei) 
because the diffractive structure functions are dominated by gluons.

Another popular approach to nuclear shadowing is the eikonal approximation.
This method is based on the assumption that the interaction between the 
fluctuation $|h_{k} \rangle$ and nucleons of the target leaves the fluctuation 
unchanged (i.e., the interaction is diagonal in the appropriate variables).
This allows the subprocess to be iterated and summed so that the 
$|h_{k} \rangle$-nucleus total cross sections, $\sigma_{h_{k} A}$, entering Eq.~(\ref{dec2}) via ${\hat \sigma}_{k,k^{\prime}}^{A}$, can be expressed in an exponential form via 
the {\it total} $\sigma_{h_{k} N}$ cross sections. This procedure is termed eikonalization,
hence the name of the approximation.

The profound conceptual difference between the leading twist and eikonal approaches is 
readily revealed by including QCD evolution. Indeed, in QCD quark-antiquark pairs and 
gluons are radiated by the partons constituting the fluctuations $|h_{k} \rangle$ before 
and during the interaction with the target, so that the fluctuations $|h_{k} \rangle$ 
necessarily mix. This is included in the leading twist approach and ignored in the eikonal 
approximation. As a result, the two considered approximations predict a different size 
and $Q^2$-dependence of the shadowing correction. For example, the longitudinal structure 
function $F_{L}^{A}$ is very sensitive to the mixing of the $q \bar{q}$ and $q \bar{q}g$ 
fluctuations of the incoming photon and thus can serve as a sensitive observable to 
distinguish between the leading twist and eikonal approximation predictions.

At the outset, it may be useful to clarify the relation of this work to
a number of works  on the small-$x$ scattering off nuclei (for nice pedagogical reviews and references to original papers, see e.g., \cite{CGC}).
We are working in a kinematic region in which the leading twist contribution 
dominates diffraction and one has to take into account logs of $Q^2$ and one 
or two logs of $x_0/x$ in the evolution of the parton distributions. 
This can be accounted for by the NLO DGLAP evolution
equations. 
The approaches of \cite{CGC} focus on a resummation of the 
logs of $x_0/x$ (cf. the BFKL approximation) which is not necessary for this 
kinematic region and neglect certain logs of $Q^2$.

This paper is organised as follows.
In section~\ref{sec:main} we review the leading twist approximation for 
nuclear shadowing and give predictions for the quark and gluon nuclear 
densities and for the nuclear structure functions $F_{2}^{A}$, 
using as input a broad range of leading twist diffractive parton 
distribution functions (DPDFs), 
which successfully describe the HERA hard diffraction data. 
We also review the eikonal approximation approach to nuclear shadowing and discuss 
various serious problems associated with it.
We dedicate section~\ref{sec:longitudinal} to the longitudinal nuclear structure
function, $F_{L}^{A}$, which is predicted to be very different within the two approaches,
and indicates general qualitative patterns of contrasting behaviour for hard processes.
Discussions and conclusions are presented in section~\ref{sec:conc}.
In appendix~\ref{appendix:a} we specify and compare the different parameterizations of the DPDFs and observe that, in spite of significant differences in the details of these, they 
all lead to fairly similar probabilities for quark-induced and gluon-induced diffraction  
and nuclear shadowing
at small $x$. 
We also observe that all of these parameterizations lead to a very large probability of leading twist diffraction in gluon-induced processes, which is comparable to the 
black-body limit of $50 \%$ for $x\leq 10^{-4}$ at $Q^2 \sim 4$~GeV$^2$.
Finally, the parameterization of the nuclear one-body density for a range of nuclei considered is given in appendix~\ref{appendix:b}.

\section{The leading twist and eikonal approaches to nuclear shadowing}
\label{sec:main}

In this section we review both the leading twist and eikonal approaches to nuclear
shadowing (subsections \ref{subsec:lt} and \ref{subsec:eik}). A critical
comparison of the approaches is presented in subsection \ref{subsec:com}.

\subsection{The leading twist approach}
\label{subsec:lt}

The basis of our present day understanding of nuclear shadowing in high-energy scattering on nuclei was formulated in the seminal work by V.~Gribov \cite{Gribov}.
The key observation was that, within the approximation that the radius of the strong interactions is
much smaller than the average inter-nucleon distances in nuclei, there
is a direct relationship between nuclear shadowing in the total hadron-{\it nucleus} cross section
and the cross section for diffractive hadron-{\it nucleon} scattering.
While the original derivation was presented for hadron-deuterium scattering, it can be straightforwardly generalized to lepton-nucleus DIS.

\subsubsection{The master formulae for leading twist nuclear shadowing}

In this subsection we aim to motivate a master formula which relates nuclear shadowing 
to diffraction off a single nucleon and includes the scattering off an arbitrary number of nucleons. The main result of the work by Frankfurt and Strikman \cite{FS99} is the observation that the Gribov theory can be generalized to calculate the leading twist component of nuclear shadowing for each nuclear parton
distribution {\it separately}. We shall review the principal steps of the
derivation below. The starting point is the generalization of Gribov's result
to DIS of leptons on deuterium.
The nuclear shadowing correction to the deuteron structure
function, $\delta F_{2}^{D}=F_{2}^{p}+F_{2}^{n}-F_{2}^{D}$,
where $F_{2}^{D}$, $F_{2}^{p}$, $F_{2}^{n}$ are the deuteron,
proton, neutron structure functions, respectively,
can be presented in the form
\begin{equation}
\delta F_{2}^{D}(x,Q^2)=2\frac{1-\eta^2}{1+\eta^2}\int dk^2_t
d x_{\Pomeron} F_{2}^{D(4)}(\beta,Q^2,x_{\Pomeron},k_{t}^2)
F_D(4k_t^2+4x_{\Pomeron}^2m_N^2) \ .
\label{deu1}
\end{equation}
Here $F_{2}^{D(4)}$ is the diffractive structure function of the
proton in terms of the usual diffractive variables, $Q^2 = -q^2$,
$\beta \equiv -q^2/(2 q \cdot k) \equiv Q^2/(Q^2+M_X^2 - k^2) \approx Q^2/(Q^2 + M_X^2)$ and
$x_{\Pomeron} \equiv 2q \cdot k/(2 q \cdot p) \equiv x /\beta \approx (Q^2 + M_X^2)/(2 p \cdot q)$; $k_{t}$ is the transverse momentum transfer to the nucleon;
$\eta$ is the ratio of the real to imaginary parts of the diffractive scattering 
amplitudes\footnote{The factor associated with a non-zero $\eta$ is absent in the 
original derivation of \cite{Gribov} since the intercept of the Pomeron
trajectory $\alpha_{\Pomeron}(0)$ was assumed to be equal to
unity, see Eq.~(\ref{eq:Migdal}).}  
for the reaction $\gamma^{\ast}+N \to X +N$; $F_D$ is
the deuteron electromagnetic form factor\footnote{For simplicity we give the expression
for the spin-less deuteron. For the spin-one deuteron the relevant form factors were
calculated in \cite{FG}.}.

The relevant Feynman graph for the forward scattering amplitude of the double scattering 
diagram is shown in Fig.~\ref{fig:deu}, which also serves to define the four-momentum flow. 
In particular, the momentum transfer to the proton $k = p' - p$ is mostly 
transverse\footnote{The transverse plane is defined relative to the virtual-photon nucleon 
center of mass frame. Given that the momentum is mostly transverse, to a very good approximation, we use $t$ and $k_t^2$ interchangeably.}, so that $k^2 \approx - k_t^2$. In order to find the shadowing 
correction to the total virtual photon-deuteron cross section (and hence $\delta F_{2}^{D}$) 
one needs to find the imaginary part of the forward amplitude  of 
the diagram in Fig.~\ref{fig:deu}.

We can also obtain Eq.~(\ref{deu1}) in a somewhat different way than
in the original Gribov paper by using the Abramovsky-Gribov-Kancheli
cutting rules \cite{AGK} .
Instead of considering directly the imaginary part
of the total rescattering amplitude, one can first consider the cross section
corresponding to the diffractive cut of the amplitude in Fig.~\ref{fig:deu} 
(see Fig.~1 of \cite{FS99}).
In this case, the answer is given by
the same diffractive cross section that enters lepton-nucleon
diffractive scattering, i.e., the diagram is proportional to $|\mbox{Im} A|^2 + |\mbox{Re} A|^2$, where $A$ is the amplitude for diffractive lepton-proton scattering. 
One then compares the
expression for the imaginary part of the forward amplitude of the double
 scattering diagram to that with the diffractive cut. Using the cutting rules,
 one observes that the former is
proportional to $|\mbox{Im} A|^2 - |\mbox{Re} A|^2$,
while  $F_{2}^{D(4)}$ is proportional to $|\mbox{Im} A|^2 + |\mbox{Re} A|^2$.
This explains how the 
answer for $\delta F_{2}^{D}$ can be expressed in terms of the nucleon diffractive
 structure functions, $F_{2}^{D(4)}$, as well as the presence of the factor
 $(1-\eta^2)/(1+\eta^2)$ in Eq.~(\ref{deu1}).

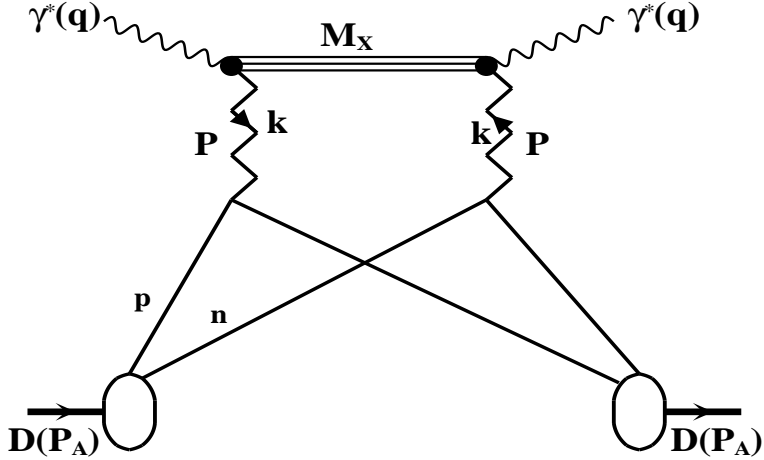
\EPSFIGURE[ht]{gr.epsi,height=6cm, width=10cm}{Feynman diagram demonstrating the connection between the nuclear shadowing correction to the deuteron structure function, $\delta F_{2}^{D}$, and the proton diffractive structure function $F_{2}^{D(4)}$.
\label{fig:deu}}

The deuteron form factor arises from the non-relativistic calculation of
the overlapping integral of deuteron wavefunctions $\psi_D$. 
Indeed, when one nucleon with a three-momentum $\vec{p}$ receives a three-momentum
$\vec{k}$=($\vec{k_t}$, $k_3$ $\approx$ $m_N x_{\Pomeron}$), 
the other nucleon with a three-momentum $-\vec{p}$ receives a three-momentum
$-\vec{k}$ (see Fig.~\ref{fig:deu}). Hence, the lower (nuclear) part of the diagram in Fig.~\ref{fig:deu} is proportional to $\int d^3p ~\psi_D(\vec{p}) ~\psi(\vec{p}+\vec{k}) = F_D (4 k^2)$ because the electromagnetic form factor is defined as $F_D(q^2) \equiv \int d^3p ~\psi_D(\vec{p}) ~\psi(\vec{p}+\vec{q}/2)$.
When the argument of the form factor is large, i.e., when 
either $x_{\Pomeron} \geq 1 /(m_N r_D)$
or $k_t \gg 1/r_D$, the damping of the form factor annihilates $\delta F_{2}^{D}$ 
($r_D$ is the average electromagnetic deuteron radius).

Equation~(\ref{deu1}) can be generalized to DIS on any nucleus. The
shadowing correction to the nuclear inclusive structure function,
which arises due to scattering on any {\it pair} of nucleons, 
$\delta F_{2}^{A(2)}$, can be written in the form
\begin{eqnarray}
&&\delta F_{2}^{A(2)}=\frac{A(A-1)}{2} 16 \pi Re \Bigg[ \frac{(1-i\eta)^2}{1+\eta^2}\int d^2b \int_{-\infty}^{\infty} dz_1 \int_{z_{1}}^{\infty} dz_2 \int_x^{x_{\Pomeron,0}}
dx_{\Pomeron} \times \nonumber \\
&&F_{2}^{D(4)}(\beta, Q^2, x_{\Pomeron}, k_t^2) \bigg|_{k_t^2=0}
\rho_A(b,z_1)\rho_A(b,z_2) e^{ix_{\Pomeron}m_N(z_1-z_2)} \Bigg] \ .
\label{deltafa}
\end{eqnarray}
Here $\rho_A$ is the nuclear one-body  density per nucleon normalized
such that $\int d^3 \vec{r} ~\rho_A (\vec{r}) = 1$; $z_1$ and $z_2$ are the longitudinal coordinates of the two nucleons involved in the scattering;
$\vec{b}$ is the impact parameter of the projectile with respect to the center of the nucleus.
 The upper limit of integration, $x_{\Pomeron,0}$ will be discussed later in the text. A non-vanishing longitudinal momentum transfer to the target is taken into account by the 
factor\footnote{This factor is analogous to the one which was derived within the vector 
meson dominance
model of photoproduction and electroproduction (for a review see \cite{Bauer78}). 
In the context of DIS it was first introduced in the second reference of \cite{AJM}.} 
$\exp(i x_{\Pomeron}m_N(z_1-z_2))$. 
The implied suppression for large $x_{\Pomeron} m_N$ 
is implicitly included in Eq.~(\ref{deu1}) by the 
$4 x_{\Pomeron}^2 m_{N}^2$ term in the argument of the deuteron form factor.
 
We determine the ratio of the real to imaginary parts of the diffractive scattering 
amplitude, $\eta$, by assuming a universal exchange and using the analytical properties 
of the scattering amplitude as a function of energy, as suggested by Gribov and Migdal 
\cite{Migdal}. Applying the Gribov-Migdal expression to our case, we determine $\eta$ 
as a function of the intercept of the Pomeron trajectory, $\alpha_{\Pomeron}(0)$,
\begin{equation}
\eta=\frac{\pi}{2} \frac{\partial \ln A}{\partial \ln 1/x}=\frac{\pi}{2}\big(\alpha_{\Pomeron}(0)-1 \big) \ .
\label{eq:Migdal}
\end{equation}

Different values for $\alpha_{\Pomeron}(0)$ obtained by various groups correspond to 
different values for $\eta$. In our analysis, we used $\eta=0.22$ in conjunction with the ACWT parameterization of DPDFs \cite{ACWT} ($\alpha_{\Pomeron} (0) = 1.14$) and $\eta=0.32$ in conjunction with the H1 \cite{H1an} parameterization ($\alpha_{\Pomeron} (0) = 1.20$). 

The integration over $k_{t}$ is absent in Eq.~({\ref{deltafa})
since it is assumed
that the slope of the elementary diffractive amplitude for the
reaction $\gamma^{\ast}+N \to X +N$ is much smaller than the slope of the nuclear
electromagnetic form factor\footnote{This is analogous to the optical model approximation
of the Glauber model which is known to work very well for $A\geq 12$.}.
This allows one to replace the diffractive nucleon structure function by its value 
at $k_t^2 = 0$.

Equations~(\ref{deu1}) and (\ref{deltafa}) include only the spin
non-flip
part of the diffractive cross section, which is proportional
to $F_{2}^{D(4)}$. Spin-flip effects in diffraction are highly
suppressed at small $t$
by the structure of the Pomeron vertex and thus
can be safely neglected.

The crucial observation made in \cite{FS99} was that Eq.~(\ref{deltafa})
can be generalized to nuclear parton densities. Indeed,
the QCD factorization theorems for inclusive DIS \cite{factorization} and
hard diffraction \cite{Collins} allow one to express the
inclusive and diffractive structure functions, entering the left and right
hand sides of Eq.~(\ref{deltafa}), respectively, as a convolution of the {\it same} hard scattering
coefficients with the corresponding, i.e., inclusive and
diffractive, parton distributions. Equating the terms in front of
each hard scattering coefficient, one immediately arrives at the key
expression for the shadowing correction to the nuclear parton
distributions, $\delta f_{j/A}^{(2)}(x,Q^2)$,
\begin{eqnarray}
&&\delta f_{j/A}^{(2)}(x,Q^2)=\frac{A(A-1)}{2} 16
\pi Re \Bigg[\frac{(1-i\eta)^2}{1+\eta^2}\int d^2b \int_{-\infty}^{\infty} dz_1 \int_{z_{1}}^{\infty} dz_2 \int_x^{x_{\Pomeron, 0}}
dx_{\Pomeron} \times \nonumber \\
&&f_{j/N}^{D}(\beta, Q^2,x_{\Pomeron},0) 
~\rho_A(b,z_1)~\rho_A(b,z_2)~e^{ix_{\Pomeron}m_N(z_1-z_2)} \Bigg] \ .
\label{deltaf}
\end{eqnarray}
Here $j$ indicates a generic parton label (i.e., a gluon, or a quark of a particular flavour); 
$f_{j/N}^{D}$ is the diffractive parton distribution function (DPDF)
of the nucleon\footnote{We do not distinguish
between protons and neutrons since we consider small-$x$ scattering
in the vacuum channel.}, for parton of type $j$.
Equation~(\ref{deltaf}) is the main result of \cite{FS99} and demonstrates 
the intimate connection between the nuclear shadowing correction to nuclear inclusive parton distribution functions (PDFs) and the DPDFs of the nucleon.  
Since $f_{j/N}^{D}$ obeys the leading twist DGLAP evolution equation, the $Q^2$-evolution of $\delta f_{j/A}^{(2)}$ is also governed by DGLAP, i.e., it is {\it by definition} a leading twist contribution. This explains why the approach of \cite{FS99} can legitimately
be called the leading twist approach.
However, the validity of Eqs.~(\ref{deu1},\ref{deltafa},\ref{deltaf})
does not require the absence of the higher twist effects in diffraction.
Hence, as soon as a particular parameterization of the diffractive cross sections
fits the data, one can use this parameterization to calculate the shadowing correction to the structure functions, using Eqs.~(\ref{deu1},\ref{deltafa},\ref{deltaf}), without
addressing the question of decomposition of the diffractive cross section over twists.

Ignoring for a moment the contribution to nuclear shadowing 
arising from the interactions with three and more nucleons of the target, 
the connection of $\delta f_{j/A}^{(2)}$ to nuclear ($f_{j/A}$) and nucleon ($f_{j/N}$) PDFs is given by
\begin{equation}
f_{j/A}=Af_{j/N}-\delta f_{j/A}^{(2)} \, , 
\label{qq}
\end{equation}
and is depicted graphically in Figs.~\ref{fig:uq} and \ref{fig:gl} for the up quark and gluon distributions, respectively.

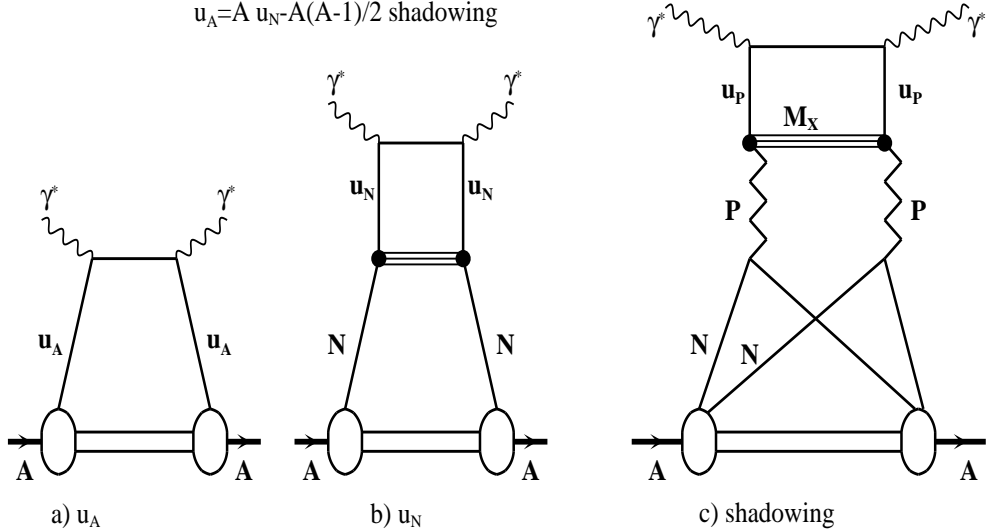
\EPSFIGURE{u.epsi,height=7cm,width=13cm}{Feynman diagrams corresponding to Eq.~(\ref{qq}) representing the nuclear quark distribution (a) as an incoherent sum of the quark distributions in the nucleons (b) minus the nuclear shadowing correction (c). Note that one needs to take the imaginary part of the forward amplitudes in order to find the structure functions.
\label{fig:uq}}

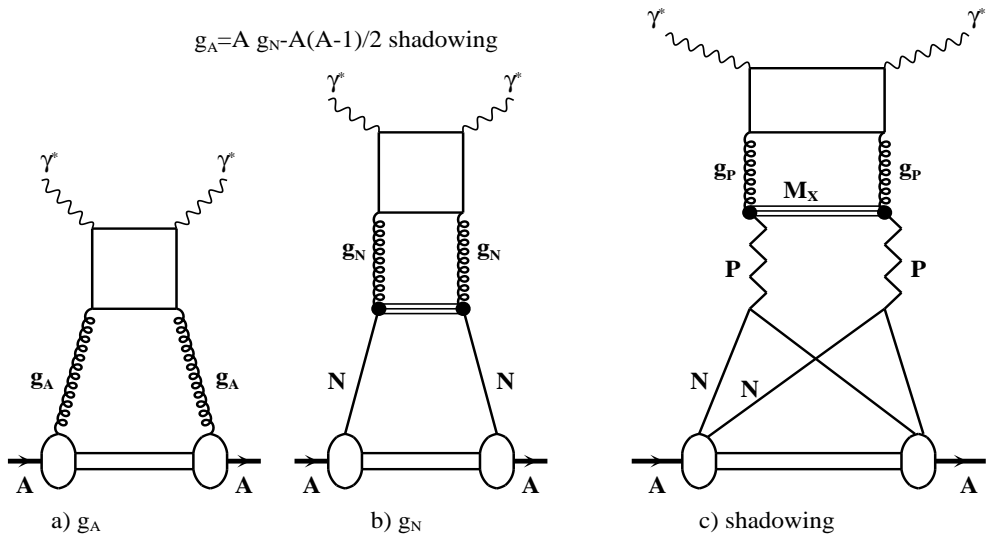
\EPSFIGURE{gl.epsi,height=7cm,width=13cm}{Feynman diagrams corresponding to Eq.~(\ref{qq}) representing the nuclear gluon distribution (a) as an incoherent sum of the gluon distributions in the nucleons (b) minus the nuclear shadowing correction (c). Note that one needs to take the imaginary part of the forward amplitudes in order to find the structure functions.\label{fig:gl}}

However, Eq.~(\ref{deltaf}) describes only one piece of the shadowing correction, namely 
the part which arises from the interaction with any two nucleons.
Since the strength of the interaction with several nucleons is large, 
for a sufficiently heavy nucleus the interaction with three and more 
nucleons plays an important role. Such an interaction cannot be 
expressed directly in terms of diffraction on the nucleon and has to 
be modelled. We shall use a scheme inspired to a certain extent by the 
quasi-eikonal approximation, which successfully describes soft total 
hadronic cross sections of high energy hadron-nucleus interactions. 
In this approximation, the amplitude for the forward virtual 
photon-nucleus scattering is given by graphs containing virtual photon-nucleon 
cross sections of two kinds. These are the cross sections of diffractive 
scattering and the production of inelastic intermediate states, and cross 
sections of ``elastic'' rescattering of the produced inelastic states.
In the formalism of cross section eigenstates \cite{Goodwalker}, 
where one introduces the notion of a distribution of cross sections for the
scattering state, the amplitude for the scattering off $k$ nucleons is
proportional to $\left<\sigma^{k}\right>$
where $\left< ... \right>$ denotes averaging over the scattering state
wavefunction. 
The quasi-eikonal approximation corresponds to the
approximation
\begin{equation}
{\left<\sigma^{k}\right> \over \left<\sigma\right>}=\sigma_{eff}^{k-1}\ .
\end{equation}
One can interpret $\sigma_{eff}$ as the average strength of the
interaction of the produced configurations.
In this approximation the total cross sections can be ``eikonalized'' or summed to
infinity in an exponential form in order to present the answer in a compact
analytical expression (cf. \cite{Bauer78}).
Introducing the full nuclear shadowing correction to $f_{j/A}$, $\delta f_{j/A}$, where
$\delta f_{j/A}=A f_{j/N}-f_{j/A}$,
the quasi-eikonal approximation for $\delta f_{j/A}$ reads
\begin{eqnarray}
&&\delta f_{j/A}(x,Q^2)=\frac{A(A-1)}{2} 16
\pi Re \Bigg[\frac{(1-i\eta)^2}{1+\eta^2} \int d^2b \int_{-\infty}^{\infty} dz_1 \int_{z_{1}}^{\infty} dz_2 \int_x^{x_{\Pomeron, 0}}
dx_{\Pomeron} \times \nonumber \\
&&f_{j/N}^{D}(\beta, Q^2,x_{\Pomeron},0) ~\rho_A(b,z_1) ~\rho_A(b,z_2)e^{ix_{\Pomeron}m_N(z_1-z_2)} e^{-(A/2)(1-i\eta) \sigma_{eff}^{j} \int^{z_{2}}_{z_{1}} dz \rho_{A}(z)} \Bigg] .
\label{qe}
\end{eqnarray}

The rescattering cross section $\sigma_{eff}^{j}$ is discussed further in appendix~\ref{appendix:a}.

Equation~(\ref{qe}) implies that the rescattering cross section for
the interaction with three and more nucleons, $\sigma_{eff}^{j}$, is
the same as the rescattering cross section for the interaction with
two nucleons. In general, this is not true since hadronic fluctuations with very 
different cross sections contribute to the total cross section of the interaction of a 
hard probe (virtual photon, $W$-boson, etc.) with a parton of the
nucleus, and this broad dispersion over cross sections should be kept in mind. 
However, for nuclei with realistic finite $A$, numerical studies with models which include 
effects associated with the distribution over cross sections demonstrate that 
deviations from the quasi-eikonal approximation are small \cite{FMS93,Harrington}. 
Similar conclusions were reached in \cite{AFS99} for the case of nuclear shadowing,
where an even broader range of possible cross section fluctuations was considered.

We used Eq.~(\ref{qe}) in order to estimate $\delta f_{j/A}$ at some initial $Q^2=Q_{0}^2$ 
($Q_{0}^2=4$ GeV$^2$ in our case). The result was used as an initial condition for the QCD 
evolution of $f_{j/A}$ to higher scales $Q^2$. We would like to stress that it would 
be incorrect to apply Eq.~(\ref{qe}) at any arbitrarily large $Q^2$: as $Q^2$ increases, the 
QCD radiation of gluons and $q \bar{q}$ pairs increases the dispersion of cross section 
fluctuations and hence makes the quasi-eikonal approximation less justified.

\subsubsection{The transition between shadowing and antishadowing: setting $x_{\Pomeron, 0}$.}

The upper limit of integration over $x_{\Pomeron}$ in
Eqs.~(\ref{deltafa}, \ref{deltaf}, \ref{qe}), $x_{\Pomeron, 0}$, deserves a
special discussion. In this subsection we motivate our choices for the numerical values of this 
parameter for the various parton species and explain that in our model this parameter 
effectively acts as the transition point from the shadowing region at low $x$ to the 
antishadowing region at larger $x$. 
In \cite{FS99}, it was assumed
that $x_{\Pomeron, 0} = 0.02$. This choice of
$x_{\Pomeron, 0}$ was motivated by two observations. Firstly,
only for $x_{\Pomeron} \leq 0.02$ is it valid to assume that
the Pomeron contribution dominates diffraction, i.e., that sub-leading
Reggeon exchanges can be safely neglected.
 Secondly, larger $x_{\Pomeron, 0}$ means that Eq.~(\ref{deltaf}) can be applied to larger Bjorken $x$. However, for $x > 0.02$, nuclear parton distributions are expected to receive an additional contribution from nuclear antishadowing (the mechanism leading to the experimentally observed enhancement of nuclear inclusive structure functions), which is not included in Eq.~(\ref{deltaf}). This makes the application of Eq.~(\ref{deltaf}) for $x > 0.02$ not well motivated.

The HERA diffractive data indicate that Pomeron exchange dominates diffraction for $x_{\Pomeron} \leq 0.01$. At higher $x_{\Pomeron}$ it becomes necessary 
to take into account sub-leading Reggeon exchanges, although 
the ``Pomeron term'' appears to be giving the dominant contribution
for $x_{\Pomeron}\leq 0.03$. At the same time the separation of the
diffractive cross section into these two contributions in this
kinematic region is definitely not unique and leads to some uncertainties.
However, the sensitivity of the predictions resulting from Eq.~(\ref{deltaf}) 
to the particular choice of $x_{\Pomeron, 0}$ is greatly reduced by the factor
associated with the nuclear wavefunction, i.e., by the suppression factor
$\rho_A(b,z_1) \rho_A(b,z_2) \exp(i x_{\Pomeron}m_N(z_1-z_2))$.

Moreover, to perform a self consistent description of the parton
densities for the whole range of $x$ one needs to incorporate effects of
enhancement of the parton densities at higher $x$. 
For the gluon such an enhancement follows from an energy-momentum sum rule
analysis of the DIS nuclear data \cite{FS88,FLS} which revealed that
the light-cone fraction carried by gluons in a nucleus and in a free nucleon
is practically the same. For the valence quarks such an
enhancement follows from the baryon charge sum rule \cite{FS88}.
For the antiquark/sea channel there are no general arguments to determine the effect
of the enhancement and in fact no enhancement is observed in the Drell-Yan 
process\footnote{We note in passing that the observed pattern of the $A$-dependence of 
nuclear parton densities 
at $x \sim 0.1 - 0.2$
may be related to the origin of nuclear forces 
\cite{FS88,Gousset96}.} \cite{DY}. Hence we followed a somewhat simplified scenario of 
\cite{FS88,FLS} which assumes that the enhancement is present only in the gluon and
valence quark channels.
The absence of the enhancement for the sea in principle allows one to take
$x_{\Pomeron, 0}$ arbitrarily large. However, in our numerical analysis we set 
$x_{\Pomeron, 0}=0.1$  for the sea. We explicitly checked that the answer is 
insensitive to the particular choice of $x_{\Pomeron, 0}$ due to the nuclear 
wavefunction suppression and the dominance of small-$x$ values for the sea quark 
distributions.

We take $x_{\Pomeron, 0}=0.03$ for the gluons.
For the gluon  case we need to introduce a transition from the shadowing to 
the antishadowing regions for some particular value of $x = x_{tr}$, which in general
should depend on $A$. Arguments were given in \cite{FS88} to the effect 
that the enhancement should be restricted to the region of $x < 0.2$, 
and could start somewhere in the range $x_{tr}  \in [0.02, 0.05]$.  
In principle we could keep the same $x_{\Pomeron, 0} = 0.1$ as for the sea
quarks and introduce an enhancement term for $x > x_{tr}$ to compensate for
the shadowing term and to ensure an enhancement.
As one can see from the structure of Eqs.~(\ref{deltafa},
 \ref{deltaf},
 \ref{qe}), the shadowing correction vanishes when $x=x_{\Pomeron,
 0}$, 
and is strongly suppressed close to it. Hence we find that for our 
numerical analysis it is simpler to take $x_{\Pomeron, 0} = x_{tr}$ 
and to neglect the enhancement term for $x \leq  x_{tr}$.

The transition point between the shadowing and antishadowing regions for gluon 
distributions in nuclei has never been measured experimentally.
The best one can do is to infer some information about $x_{tr}$ for the gluons 
indirectly, using QCD evolution.
Our choice of $x_{\Pomeron, 0}$ for the gluon distributions was motivated by the analysis in 
\cite{Gousset96} of the NMC high statistics data on the
ratio of structure functions $F_{2}$ of Tin ($^{118}$Sn) and Carbon ($^{12}$C) \cite{NMC-Sn-C}.
This analysis indicated that the transition between shadowing and
antishadowing occurs for $0.023 \leq x \leq 0.035$ and $Q^2 \approx 4$~GeV$^2$. 
Thus, we use $x_{\Pomeron, 0}=0.03$ for the gluon distributions in our analysis. 
However, a choice of $x_{\Pomeron, 0}$ does not fix entirely the shape of the gluon 
distribution $g_{A}(x)$. We assumed a quadratic polynomial form for the antishadowing 
correction to $g_{A}(x)$ on the interval $0.03 \leq x \leq 0.2$. The three free
coefficients of the quadratic fit were chosen such that the antishadowing correction 
vanishes at $x=0.03$ and $x=0.2$ and that the momentum fraction carried by gluons 
bound in nuclei is the same as that in the free nucleon \cite{FLS}. 
A similar shape of the resulting gluon enhancement was assumed in \cite{FLS,Eskola}.

It is important to mention that, in this work, we make predictions for the shadowing 
corrections to the gluon and sea quark distributions in nuclei and do not concern
ourselves with the valence quarks. In our numerical analysis,
we keep only the Pomeron exchange contribution to $F_{2}^{D(4)}$,
which has vacuum quantum numbers. Thus, nuclear shadowing for
the valence quarks in nuclei, which is associated
with the Reggeon exchanges with non-vacuum quantum
numbers, is not considered and effects of the valence quark enhancement for $x\sim 0.1$
are neglected. 
In general, a complete model for the leading twist parton distributions in
nuclei would require conservation of the baryon number and
momentum sum rules \cite{AJM} in order to reconstruct the nuclear enhancement 
of gluon and sea quark parton distributions at moderate $x$.

For the gluon and sea quark channels the
corrections due to the $A$-dependence of the valence quark distributions
(which are anyway rather small for the $x$-range we discuss)
enter only through $Q^2$-evolution and are very small, see e.g., \cite{FLS,Eskola}.

\subsubsection{Numerical predictions for nuclear PDFs and $F_2^A$}

In this subsection, we briefly discuss the DPDFs that we used in our analysis 
(we defer a more detailed specification to appendix~\ref{appendix:a}). We then 
present predictions for nuclear shadowing, 
both on the parton level (for both sea quarks and gluons) and for the 
nuclear structure functions, $F_2^A$, for a range of $Q^2$ and $A$.

High precision data on diffractive processes, taken at HERA, and
their QCD analysis allow the diffractive parton distributions of
the proton to be determined with good accuracy. In particular,
both the ZEUS \cite{ZEUS1,ZEUS2,ZEUS3} and H1 \cite{H1} collaborations confirmed that a successful
fit to the experimental data required a large gluon contribution
(the same conclusion is drawn from an analysis based on
the latest H1 data \cite{H1iechep} on inclusive diffraction).
One should especially note that the recent H1 data on diffractive
photo- and electroproduction of dijets are dominated by the gluon diffractive structure
functions \cite{H1dijet}. We consider four different parameterizations of $f_{j/N}^{D}$: fit D
of Alvero, Collins, Terron and Whitmore \cite{ACWT}; fit D of \cite{ACWT}, which has been
improved to include a low-$\beta$ contribution; the parameterization from the theoretical
light-cone model of Hautmann, Kunszt and Soper \cite{HKS}, and the H1 fit \cite{H1an} to their
own data \cite{H1}. Details of each of the parameterizations can be found in appendix~\ref{appendix:a}.
        
Predictions for nuclear parton distributions $f_{j/A}$ made with these
parameterizations of $f_{j/N}^{D}$ are fairly consistent with each other. 
The spread of the predictions reflects the theoretical uncertainty in determining 
$f_{j/A}$ in terms of $f_{j/N}^{D}$ due to experimental errors, certain differences 
between H1 and ZEUS results, and the limited $x$ and $x_{\Pomeron}$ ranges of the data.
Furthermore, \cite{ACWT} did not include the most recent ZEUS and H1 data, 
while \cite{HKS} actually did not perform a detailed fit to the data. 
In particular, the recent H1 data \cite{H1iechep}, which are most sensitive to the gluon 
diffractive structure functions, are best described by the H1 fits \cite{H1an} 
to their earlier data, while the fit of Alvero {\it et al.} \cite{ACWT} 
somewhat overestimates the dijet production rate
in the kinematics where the gluon DPDF at large $\beta$ gives the dominant contribution\footnote{This is despite the fact that it has less shadowing. This effect is a peculiarity of the choices made by the various groups when choosing parameterizations for the input diffractive quark and gluon distributions
in the regions of large ($\beta \sim 1$) and small $\beta$
 and reflects the residual uncertainties in the DPDFs.
A larger DPDF at large $\beta$ leads to a faster onset of nuclear shadowing at $x \sim 10^{-2}$, while the small-$\beta$ region determines the amount of shadowing at $x \le 10^{-4}$.}.

For nuclear shadowing of the gluon distribution in nuclei, there is also an 
important uncertainty associated with the unknown $t$-slope of the  
diffractive gluon distribution of the nucleon, $f_{g/N}^{D}$. A detailed analysis, 
presented in appendix~\ref{appendix:a}, demonstrates that 
Eqs.~(\ref{deltafa}, \ref{deltaf}, \ref{qe}) require the diffractive parton distribution 
evaluated at $t \approx 0$. This implies that the answer for the shadowing 
correction is proportional to the $t$-slope of this distribution.
The $t$-slope for the quark-dominated diffractive distributions in the proton, $B_{q}$, 
was measured in inclusive diffraction by ZEUS \cite{ZEUS2} to be $B_{q} = 7.2 \pm 1.1$~GeV$^{-2}$ (see appendix~\ref{appendix:a} for explanations on our use of $B_{q}$).
No experimental information about the $t$-slope of the gluon diffractive distribution, 
$B_{g}$, is available.
In most of the phenomenological fits it is assumed that $B_{g}=B_{q}$. 
Since the amount of shadowing in the gluon channel increases with an increase
of $B_{g}$ (for fixed total gluon-induced diffraction) we 
take a conservative attitude and consider a minimum value for the slope,  
which originates solely from the coupling of a small object to a nucleon via two gluons. 
To estimate this {\it minimal} slope we used the recent data on the 
diffractive electroproduction of $J/\psi$ by H1 \cite{H1Jpsi} which can be parameterized in 
the following $x$-dependent form
\begin{equation}
B_{g}=\Big(B_{0}+2\alpha'_{J/\psi} \ln(10^{-3}/x)\Big) \quad {\rm GeV}^{-2} \ ,
\label{eq:bg}
\end{equation}
with $B_{0} = 4.5$~GeV$^{-2}$ and  $\alpha'_{J/\psi} = 0.125$~GeV$^{-2}$.  
The slow $x$-dependence describes an increase of the slope with increasing 
energy (``shrinkage''). In order to reflect the uncertainty associated with the slope
$B_{g}$, we also considered a scenario of nuclear shadowing with $B_g = B_q$. 
For this choice, relative to that of Eq.~(\ref{eq:bg}), the shadowing of gluons, given 
by $1-g_{A}/(A g_{N})$, increases by a factor of 1.6 at $x=10^{-3}$, and of 1.4 at $x=10^{-5}$, 
for light nuclei, and by a factor of 1.4 at $x=10^{-3}$, and of 1.2 at $x=10^{-5}$, for heavy nuclei.

It is important to note that all of the parameterizations of the gluon
DPDF in the region of $x \sim 10^{-4}-10^{-5}$
reach the unitarity limit for $Q^2 \sim 4$~GeV$^2$ (see appendix~\ref{appendix:a}).
This means that the probability of diffraction in gluon-induced DIS processes exceeds 
one half for these $Q^2$. Hence, we are forced to tame the gluon DPDF at very small 
$x$ in order to comply with the unitarity constraints.
Note that, in the black-body limit, the cross section fluctuations are strongly
suppressed:  in principle reduces uncertainties in the predictions of shadowing
for these $x$, within the framework of our calculation\footnote{However, since the 
approach to this limit is not understood and must involve new effects, our confidence 
in the predictions decreases.}.

Figs.~\ref{fig:urat} and \ref{fig:glrat} show our predictions
for the ratio $f_{j/A}/(A f_{j/N})$ for the up-quark sea and
gluon distributions, respectively, in a range of nuclei.
They clearly illustrate that the effect of nuclear shadowing for gluons 
is larger than for the quarks, especially at the input scale $Q=2$ GeV. 
For instance, using the H1 parameterization, one finds at $Q=2$ GeV and 
$x=10^{-3}$ that the gluons are shadowed more than the quarks by a factor of 
1.42 for $^{12}$C and by a factor of 1.24 for $^{208}$Pb.
At $x=10^{-5}$, 
due to the increase of $B_{q}$ with energy
and the unitarity taming of the gluon mentioned previously, 
shadowing for gluons and quarks are very similar.
When the ACWT parameterization is employed, 
the gluons are shadowed more than the quarks by a factor of 
1.70 for $^{12}$C and by a factor of 1.39 for $^{208}$Pb at $x=10^{-3}$.
At $x=10^{-5}$, the factors are 1.29 for $^{12}$C and 1.11 for $^{208}$Pb.

Of the four parameterizations of diffractive parton distributions discussed 
in appendix~\ref{appendix:a}, we demonstrate only the two representative 
examples which give the largest (H1 \cite{H1an}) and smallest 
(ACWT \cite{ACWT}) nuclear shadowing effects at $x=10^{-5}$. 
This illustrates the spread of theoretical predictions for 
$f_{j/A}/(A f_{j/N})$ which arises from uncertainties in the diffractive parton 
distributions. The results are presented for $Q=2$ GeV (solid curves), $Q=5$ GeV 
(dashed curves) and $Q=10$ GeV (dotted curves). At $Q=2$ GeV, Eq.~(\ref{qe}) was used. 
For the inclusive parton distributions of the nucleon, the recent CTEQ5M 
parameterization \cite{CTEQ5} was employed. We used the nuclear PDFs, $f_{j/A}$, 
at $Q=2$~GeV as input to the DGLAP evolution equations in order to find $f_{j/A}$ at 
$Q = 5, 10$ GeV.

One can see from the figures that differences between the predictions
are large for $x \sim 0.01$, where contribution of the large 
$\beta ~\gsim ~0.3 - 0.6$ 
region dominates. 
This $\beta$ region was not strongly constrained by the 
older data from which the fits are derived.

For the region of $x\sim 10^{-3}$ predictions for sea quark 
shadowing exhibit minimal spread because the integrals 
over $x_{\Pomeron}$ in Eq.~(\ref{qe}) are dominated by regions where 
the diffractive data is of high precision. 
The spread of predictions increases again for $x \leq 10^{-4}$ reflecting  
the extrapolation of the DPDFs to the region of very high energies 
(beyond the HERA range).
In the gluon case the spread is even larger, reflecting
greater uncertainties in extraction of gluon diffractive 
distributions from the scaling violation of the $F_2^{D(3)}$ data.
Of two parameterizations presented here, only the H1 parameterization
fits the dijet data, which indicates that it may be more realistic.

\EPSFIGURE{urat3newww.epsi,height=15cm,width=15cm}{
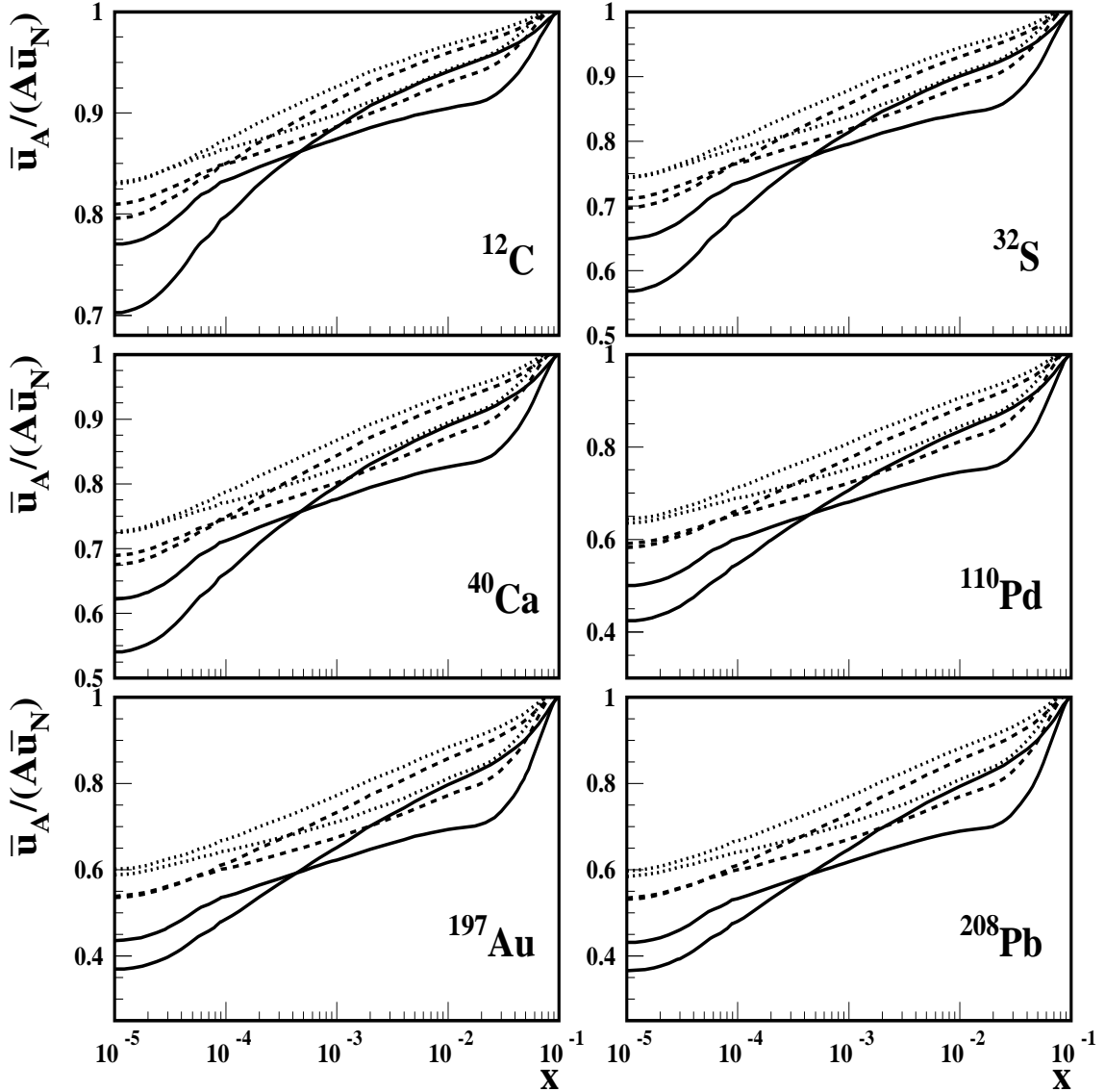
The ratio of nuclear to nucleon sea up-quark distributions,
$\bar{u}_{A}/(A \bar{u}_{N})$, at $Q=2$ GeV (solid line), $Q=5$ GeV
(dashed line), and $Q=10$ GeV (dotted line). For each case, the
scenarios with the largest (based on \protect\cite{H1an}) and smallest 
(based on \protect\cite{ACWT}) nuclear shadowing for $x=10^{-4}$ 
are presented. \label{fig:urat}}

\EPSFIGURE{glrat3newww.epsi,height=15cm,width=15cm}{
The ratio of nuclear to nucleon gluon distributions, $g_{A}/(A g_{N})$
at $Q = 2$ GeV (solid line), $Q = 5$ GeV (dashed line), and $Q = 10$ GeV
(dotted line). For each case, the scenarios with the largest 
(based on \protect\cite{H1an}) and smallest (based on \protect\cite{ACWT}) 
nuclear shadowing for $x=10^{-4}$ are presented. \label{fig:glrat}}

It was observed in \cite{FS99} that shadowing for gluons, calculated using 
the DPDFs of Alvero {\it et al.} \cite{ACWT}, is much larger than for the sea quarks. 
The present analysis demonstrates that this is quite sensitive to the particular 
choice of the input diffractive parton distribution functions. However, it would 
appear from Figs.~\ref{fig:urat} and \ref{fig:glrat} that for the smallest $x$ 
values nuclear shadowing for sea quarks is not so much different from that for the gluons.
There are several reasons for this. Firstly, and
most importantly, our numerical calculations were performed using
the slope $B_{g}$, given by Eq.~(\ref{eq:bg}), which is smaller
than $B_{g}=B_{q}$, as implemented in \cite{FS99}. Secondly, the
effective cross section for the gluon channel, $\sigma_{eff}^{g}$,
which defines the size of nuclear shadowing in Eqs.~(\ref{deltafa})
and (\ref{deltaf}) for gluons, cannot be too large, otherwise
the diffractive cross section becomes larger than allowed by
unitarity of the scattering operator.
Thus, in our model shadowing for the gluons reaches a plateau at $Q \sim 2$ GeV
for 
$x < 10^{-4}$, when the H1 parameterization is used. In case of the ACWT fit, no taming of $\sigma_{eff}^{g}$ is necessary down to $x=10^{-5}$.
 This shows that it is impossible to make realistic 
calculations of nuclear shadowing at small enough $x,Q$ without invoking the unitarity 
restrictions. Also, the reverse statement is true: estimates of the kinematical 
boundaries of the region in which unitarity restrictions play a role 
require that nuclear shadowing effects be properly taken into account!

In order to demonstrate the importance of the value of $B_{g}$, we also performed 
an analysis with larger $B_{g}$: $B_{g} = B_{q}$.
The results are presented in Fig.~\ref{fig:glrat2} for the lightest ($^{12}$C) and
heaviest ($^{208}$Pb) nuclei, using the H1 DPDF. 
The thin curves are those from Fig.~\ref{fig:glrat}
(using $B_{g}$ from Eq.~(\ref{eq:bg})), while the thick curves are obtained using
$B_{g}=B_{q}$. Again, the solid, dashed and dotted curves correspond to
$Q = 2, 5, 10$~GeV. One observes that increasing the $t$-slope of the gluon DPDF 
increases the shadowing correction to $g_{A}$, given by $1-g_{A}/(A g_{N})$.
For example, at $x=10^{-3}$ and $Q = 2, 5, 10$~GeV,  $1-g_{A}/(A g_{N})$ increases 
by $60 \%$ for  $^{12}$C and $40 \%$ for $^{208}$Pb. At $x=10^{-5}$, the increase is 
about $40 \%$ for $^{12}$C and $20 \%$ for $^{208}$Pb.

\EPSFIGURE{glrat3bsnewww.epsi,height=13cm,width=15cm}{
The ratio of nuclear to nucleon gluon distributions, $g_{A}/(A g_{N})$ at $Q=2$ GeV (solid line), $Q=5$ GeV (dashed line), and $Q=10$ GeV (dotted line). The H1 parameterization \protect\cite{H1an} for diffractive parton distributions is employed. The thin curves correspond to $B_{g}$ of Eq.~(\ref{eq:bg}), the thick curves correspond to $B_{g}=B_{q}$.
\label{fig:glrat2}}

Using our results for the sea quark ratio $\bar{q}_{A}/(A \bar{q}_{N})$
and the gluon ratio $g_{A}/(Ag_{N})$, one can readily predict the ratio of the inclusive structure functions $F_{2}^{A}/(A F_{2}^{N})$.
In next to leading order accuracy, $F_{2}^{A}$ can be presented
 in the standard form 
\cite{FP82}
\begin{eqnarray}
&&F_{2}^{A}(x,Q^2)  =  
\sum_{i} e_{i}^2\big(q^{i}_{A}(x, Q^2) + \bar{q}^{i}_{A}(x, Q^2)\big)  \label{eq:qcd} \\
& + & \frac{\alpha_{s} (Q^2) }{2 \pi} x \int^{1}_{x} \frac{dy}{y} \, \left[ C_{2}^{q} (x/y) \sum_i e_{i}^2\big(q^{i}_{A}(y,Q^2) + \bar{q}^{i}_{A} (y,Q^2)\big) + \langle e^2 \rangle C_{2}^{g}(x/y) g_{A}(y, Q^2) \right] \, . \nonumber
\end{eqnarray}
Here $e_{i}$ is the electric charge of the quark with flavor $i$; $\langle e^2 \rangle=(\sum_{i} e_{i}^2)/n_f$;  $C_{2}^{q}$ and $C_{2}^{g}$ are the target-independent hard scattering coefficients
\begin{eqnarray}
&&C_{2}^{q}=\frac{4}{3} \Big[\frac{1+x^2}{1-x}\Big(\ln (\frac{1-x}{x})-\frac{3}{4}\Big)+\frac{1}{4}(9+5x)\Big]_{+} \ , \nonumber\\
&&C_{2}^{g}=n_{f} \Big((x^2+(1-x)^2)\ln (\frac{1-x}{x})-1+8x(1-x)\Big) \ , 
\end{eqnarray}
where $n_{f}$ is the number of active quark flavors and $[ \dots ]_{+}$ denotes the plus-regularization. In our analysis, $n_{f}=4$. In order to obtain $F_{2}^{N}$, one needs to replace the nuclear parton distributions by the nucleon ones in Eq.~(\ref{eq:qcd}).

We considered two scenarios of nuclear shadowing, corresponding to ACWT and H1 parameterizations of DPDFs. In the former case, since the DPDFs of the strange quarks is zero at the initial scale $Q_{0}=2$ GeV, only up and down quarks and gluons are shadowed at the initial scale.
However, when the H1 parameterization is used, $f^{D}_{u/N}=f^{D}_{d/N}=f^{D}_{s/N}$ at the initial scale so that up, down and strange quarks are shadowed.
For both of the parameterizations of DPDFs we considered, the charm quarks are not affected by the nuclear medium at the initial scale, i.e., $c_{A}=A c_{N}$. However, as a result of the QCD evolution, gluons convert into 
$s \bar{s}$ and 
$c \bar{c}$ pairs, which leads to a significant deviation of the ratios 
$s_{A}/(A s_{N})$ and 
$c_{A}/(A c_{N})$ from unity (see Fig.~\ref{fig:charm}). 

The results of the calculation using Eq.~(\ref{eq:qcd}) are presented in Fig.~\ref{fig:F2A}. In this analysis we assumed that valence quarks are shadowed the same as the sea quarks. A comparison of Figs.~\ref{fig:urat} and \ref{fig:F2A} reveals that while at $Q=5,10$ GeV, the ratios $\bar{u}_{A}/(A \bar{u}_{N})$ and $F_{2}^{A}/(A F_{2}^{N})$ are quite similar, $F_{2}^{A}/(A F_{2}^{N})$ is smaller than $\bar{u}_{A}/(A \bar{u}_{N})$ at $Q=2$ GeV due to the important contribution of the unshadowed charm quarks.

\EPSFIGURE{f2rat2w.epsi,height=15cm,width=15cm}{
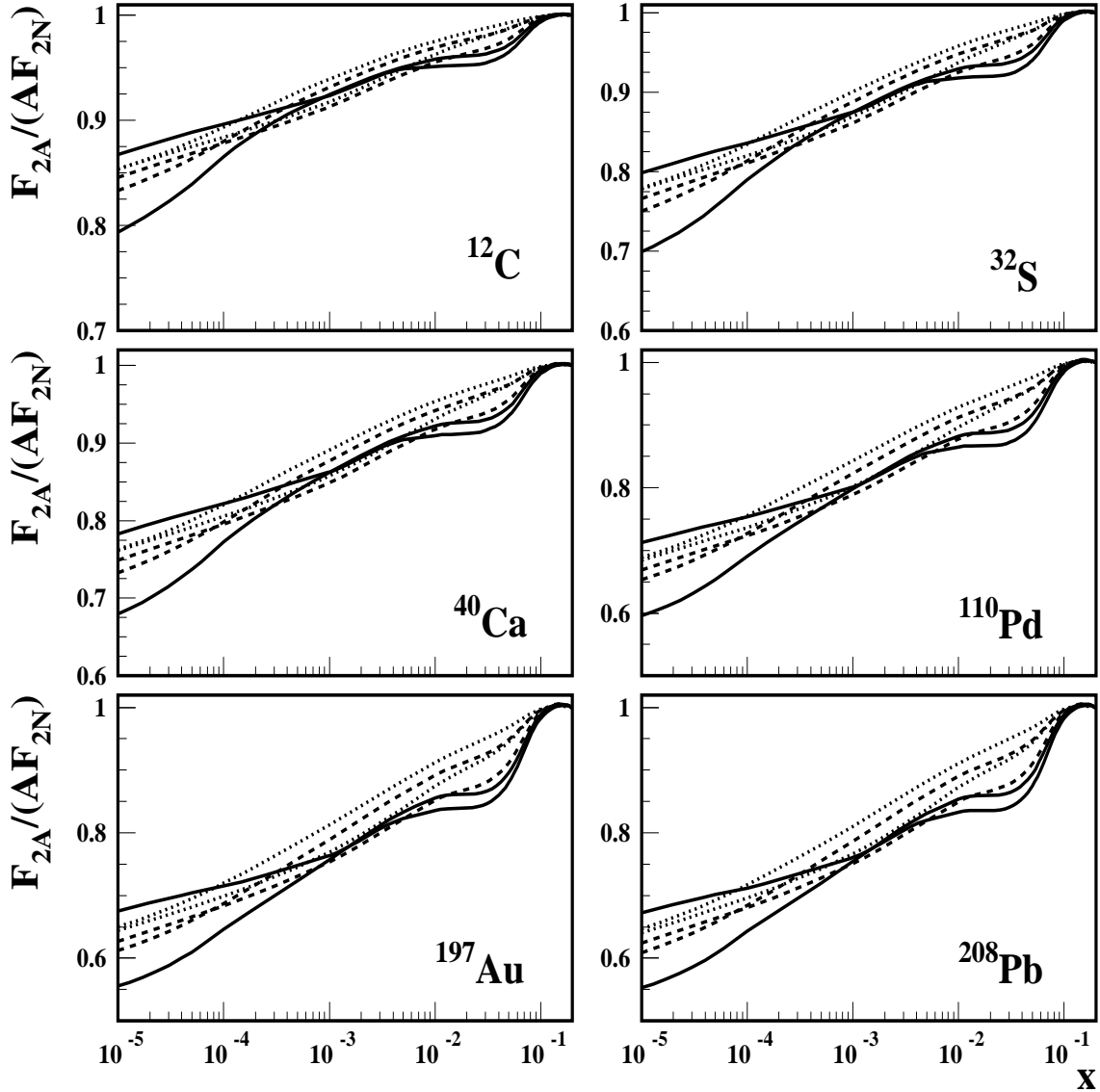
The ratio of nuclear to nucleon inclusive structure functions $F_{2}^{A}/(A F_{2}^{N})$. The solid, dashed and dotted curves correspond to  $Q = 2,5,10$~GeV and the scenarios of nuclear shadowing are as in Fig.~\ref{fig:urat}.
\label{fig:F2A}}

One can see from Fig.~\ref{fig:F2A} that the solid curves corresponding to $Q=2$ GeV run close to each other between $x=10^{-4}$ and $x=10^{-2}$. In order to understand this pattern, one should recall Eq.~(\ref{qe}) which demonstrates that $\delta F_{2}^{A}$ is linearly proportional to $F_{2}^{D(4)}$ in the low nuclear density limit. Since both the ACWT and H1 parametrizations give a good fit to $F_{2}^{D(4)}$, it is no surprise that the resulting shadowing corrections to $F_{2}^{A}$ in the appropriate range of $x$ are virtually the same for both fits.

A peculiar feature of Fig.~\ref{fig:F2A} is that the upper solid curves, which correspond to the calculation with the ACWT DPDF at $Q=2$ GeV, 
in the small-$x$ region
lie above (i.e., correspond to smaller nuclear shadowing)  all other curves corresponding to higher $Q^2$. 
There are two reasons for this. Firstly and most importantly, the absence of nuclear shadowing for charm quarks at the initial scale significantly decreases the nuclear shadowing correction to $F_{2}^{A}$. Secondly, the next to leading order contribution to  $F_{2}^{A}$ (terms proportional to $\alpha_{s}(Q^2)$ in Eq.~(\ref{eq:qcd})) has reduced shadowing as a result of the integration over $y$.

We would like to point out  that there is no reason to assume that nuclear shadowing for the charm quarks is zero at the initial evolution scale. We propose the following simple model. At small values of Bjorken $x$ charm quarks are mostly produced by QCD evolution because of the splitting $g \to c \bar{c}$. Thus, nuclear shadowing for the charm quarks at some $x$ and $Q=Q_{0}$ originates from nuclear shadowing of gluons at larger $x$ and $Q_{eff}$. We choose
\begin{equation}
\frac{c_{A}(x,Q_{0}^2)}{A c_{N}(x,Q_{0}^2)}=\frac{g_{A}(2x,Q_{eff}^2)}{A g_{N}(2x,Q_{eff}^2)} \ ,
\label{shc}
\end{equation}
where $Q^2_{eff}= 4m_c^2+Q_{0}^2=11$ GeV$^2$. Note that the gluon distributions are taken at twice the $x$-values of the charm quarks.
 
With this model for nuclear shadowing for charm quarks we re-evaluated the ratio $F_{2}^{A}/(A F_{2}^{N})$ using the H1 parametrization of DPDFs. The results are presented in Fig.~\ref{fig:F2Ach} as thin curves and compared to the corresponding curves from Fig.~\ref{fig:F2A} (thick curves). One can see that by introducing nuclear shadowing for the charm quarks at the initial scale by Eq.~(\ref{shc}),  one increases nuclear shadowing correction to  $F_{2}^{A}$ and makes it closer to the shadowing correction to $\bar{u}_{A}$. However, the size of nuclear shadowing introduced by our physically motivated model of Eq.~(\ref{shc}) is not sufficient to have 
$F_{2}^{A}/(A F_{2}^{N}) \approx \bar{u}_{A}/(A u_{N})$ at small $x$.

\EPSFIGURE{f2rat2chw.epsi,height=13cm,width=15cm}{
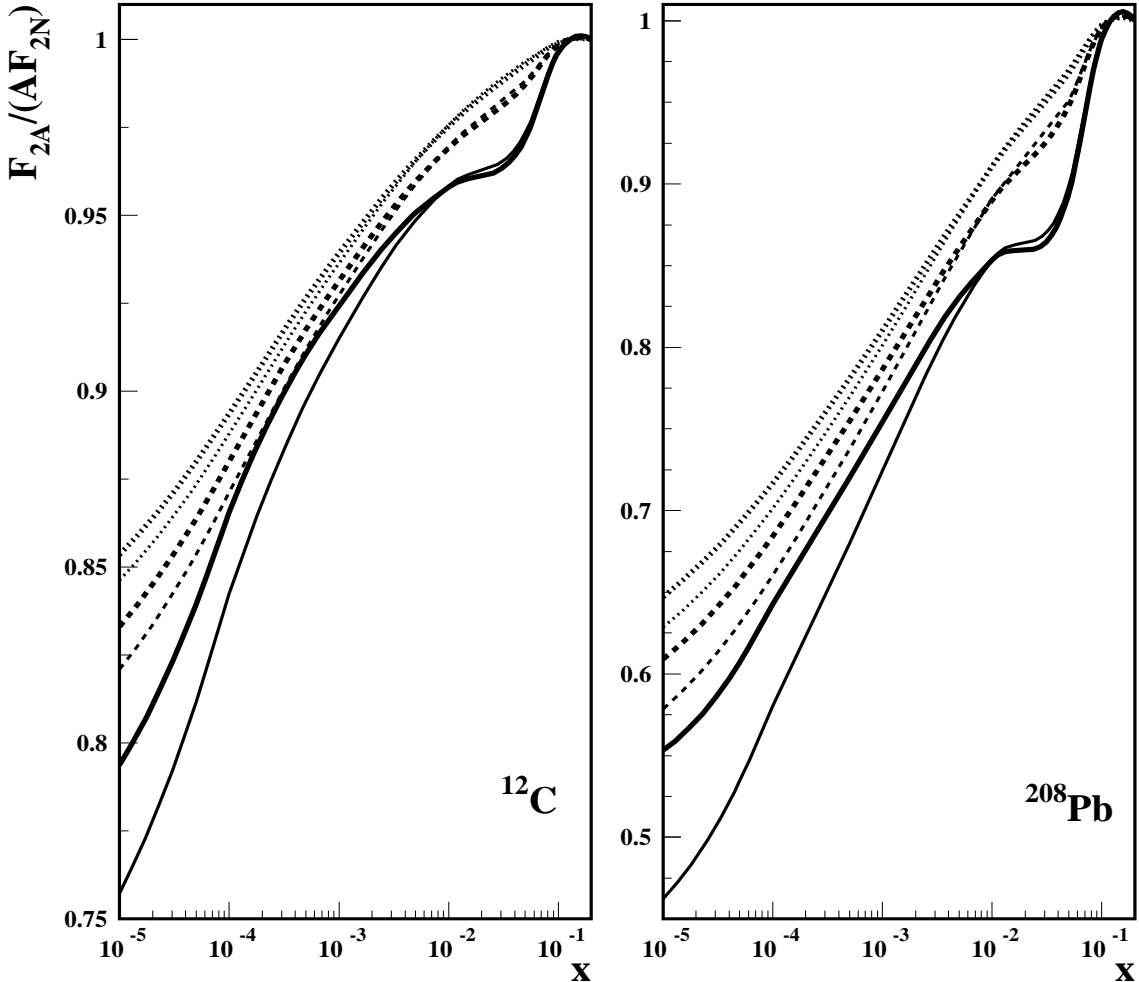
The ratio of nuclear to nucleon inclusive structure functions $F_{2}^{A}/(A F_{2}^{N})$. The thick curves are those from Fig.~\ref{fig:F2A}, while the thin ones correspond to the calculation with shadowing for the charm quarks given by Eq.~(\ref{shc}) at the initial scale $Q=2$  GeV. The calculation is done using the H1 parametrization of DPDFs.
\label{fig:F2Ach}}

Both scenarios of nuclear shadowing for  charm quarks at $Q=2$ GeV and the results of their QCD evolution are presented in Fig.~\ref{fig:charm}
 using the H1 fit to DPDFs. The thick curves represent our standard scenario without shadowing for the charm quarks at the initial scale, while the thin curves correspond to the calculation with shadowing for the charm quarks, see Eq.~(\ref{shc}). The $Q^2$-dependence of the curves is the same as in Fig.~\ref{fig:urat}.
Two features of Fig.~\ref{fig:charm} are of interest: significant antishadowing and ``wrong'' $Q^2$ dependence of nuclear shadowing -- nuclear shadowing increases as $Q^2$ increases! Both of these features can be understood as a consequence of the fact that at low $x$, charm quarks are predominantly produced by photon-gluon fusion 
$\gamma^{\ast} g \to c \bar{c}$.

\EPSFIGURE{charmw.epsi,height=13cm,width=15cm}{
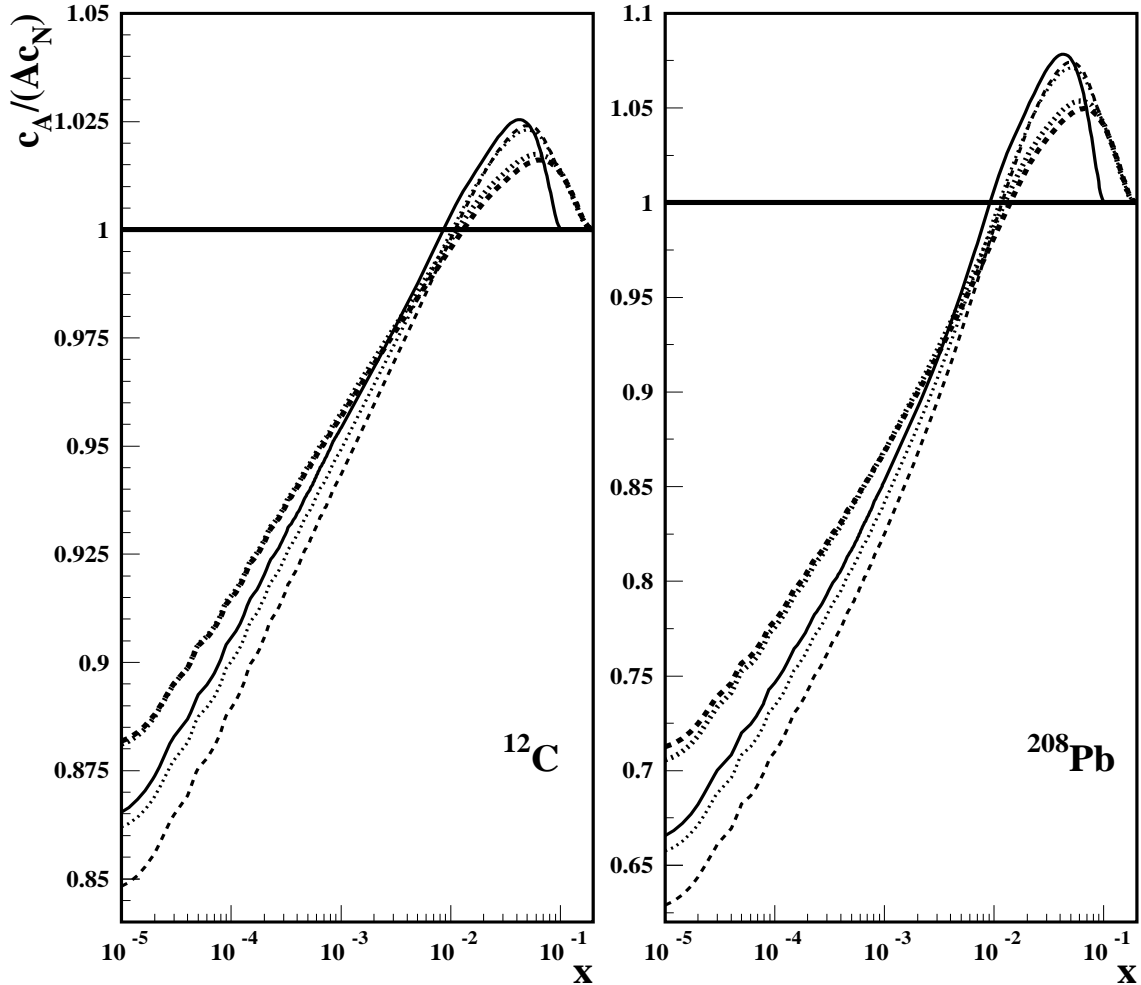
The ratio of nuclear to nucleon charm distribution functions, $c_{A}/(A c_{N})$.
The solid, dashed and dotted curves correspond to  $Q = 2,5,10$~GeV and the scenarios of nuclear shadowing are as in Fig.~\ref{fig:urat}. The calculation is done using the H1 parametrization of DPDFs.
\label{fig:charm}}

The consistent inclusion of heavy quarks in input boundary 
conditions and in QCD evolution is known to be highly non-trivial, 
and involves a certain dependence on the scheme one is using for 
the number of active quark flavours (see \cite{thorne} and references therein).
This is one reason why such a low starting scale (i.e., below the charm mass) 
is often chosen, e.g., by CTEQ, 
so that heavy quarks are only generated explicitly via photon-gluon 
fusion at the input scale. Taken literally this implies that there can 
be no shadowing of charm at the input scale since there is no 
intrinsic charm at this scale ! We attempt to correct for this, 
in a phenomenological way, by using Eq.~(\ref{shc}) above. However, 
a more accurate treatment, along the lines of variable flavour number 
scheme, would be desirable but is beyond the scope of this paper.

\subsubsection{Predictions for central collisions}

The effect of nuclear shadowing becomes even more dramatic if one
considers DIS from nuclei at small impact parameters (i.e., ``central'' collisions). 
In this case, since the density of nucleons is larger in the centre of the nucleus
than at its periphery, the number of scatterers effectively increases,
which leads to an increase of nuclear shadowing. Experimentally, one
can tag such small impact parameter events by measuring slow
``knock-out'' neutrons \cite{STZH}
which, in collider experiments with symmetric beam energies, 
can be detected with a very high efficiency ($\sim
100 \%$) by low angle neutron calorimeters.

Let us introduce a simple model for 
impact parameter-dependent nuclear PDFs, 
$f_{j/A}(x,Q^2,b^2)$.
By our definition,  $f_{j/A}(x,Q^2,b^2)$ are simply related to the usual, 
or impact parameter  integrated 
parton distributions, $f_{j/A}(x,Q^2)$, via
\begin{equation}
\int d^2 b f_{j/A}(x,Q^2,b^2) = f_{j/A}(x,Q^2) \ .
\label{normb}
\end{equation}
In common with $f_{j/A}(x,Q^2)$, $f_{j/A}(x,Q^2,b^2)$ can be presented as a difference between the impulse approximation and shadowing contributions:
\begin{equation}
f_{j/A}(x,Q^2,b^2)=\Bigg(\int^{\infty}_{-\infty} dz \rho_{A}(b,z)\Bigg) A f_{j/N}(x,Q^2)-\delta f_{j/A}(x,Q^2,b^2) \ .
\end{equation}
Here the factor $\int^{\infty}_{-\infty} dz \rho_{A}(b,z)$ guarantees the correct normalization of $f_{j/A}(x,Q^2,b^2)$ (see Eq.~(\ref{normb})). The shadowing correction $\delta f_{j/A}(x,Q^2,b^2)$ can be readily found from Eq.~(\ref{qe}) by removing the integration over $b$:
\begin{eqnarray}
&&\delta f_{j/A}(x,Q^2,b^2)=\frac{A(A-1)}{2} 16
\pi Re\Bigg[\frac{(1-i\eta)^2}{1+\eta^2} \int_{-\infty}^{\infty} dz_1 \int_{z_{1}}^{\infty} dz_2 \int_x^{x_{\Pomeron, 0}}
dx_{\Pomeron} \times \nonumber \\
&&\! f_{j/N}^{D}(\beta, Q^2,x_{\Pomeron},0) ~\rho_A(b,z_1) ~\rho_A(b,z_2) e^{ix_{\Pomeron}m_N(z_1-z_2)} e^{-(A/2)(1-i\eta)\sigma_{eff}^{j} \int^{z_{2}}_{z_{1}} dz \rho_{A}(z)} \Bigg] \ .
\label{central}
\end{eqnarray}
\EPSFIGURE{glcentrnewww.epsi,height=13cm,width=15cm}{
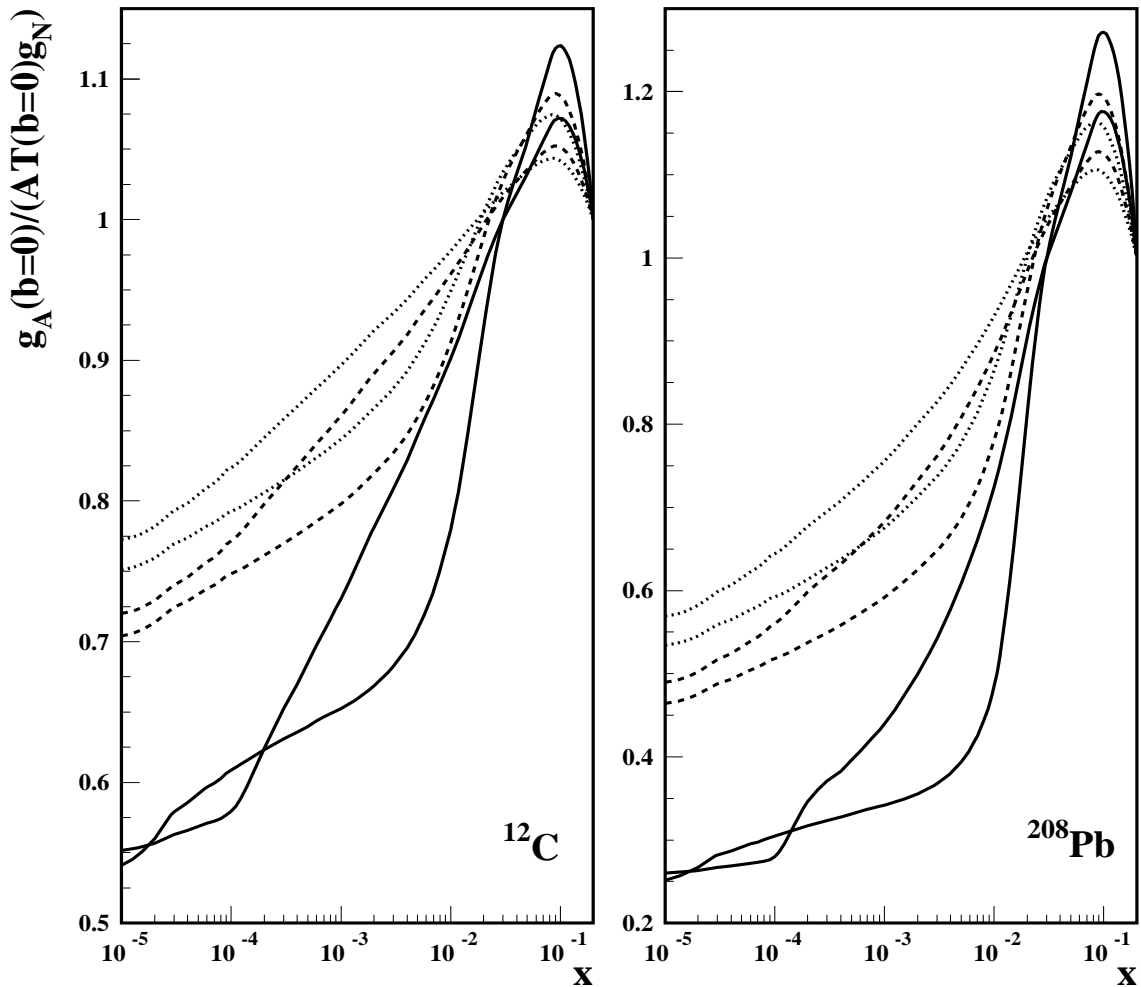
The ratio of the nuclear gluon distribution
at the zero impact parameter to that of the nucleon, $g_{A}(x,Q^2,b=0)/(AT(0)g_{N})$ at $Q=2$ GeV (solid line), $Q=5$ GeV (dashed line), and $Q=10$ GeV (dotted line). The ACWT \protect\cite{ACWT} and H1 \protect\cite{H1an} parameterizations for diffractive parton distributions are employed.
\label{fig:glrat3}}

We used Eq.~(\ref{central}) in exactly the same way that
 we used Eq.~(\ref{qe}). 
Firstly, the nuclear shadowing correction to the gluon and sea quark parton impact 
parameter-dependent parton distributions was evaluated using Eq.~(\ref{central}) 
at the initial scale $Q_{0} = 2$~GeV. Secondly, antishadowing was modelled so that 
gluons in nuclei carry the same momentum per nucleon as in the free nucleon. 
Thirdly, the shadowing and antishadowing calculations defined the input to the QCD 
evolution to higher $Q^2$ scales. The results of such an analysis for the gluon 
distributions at the zero impact parameter are presented in terms of the ratio
$g_{A}(x,Q^2,b=0)/(A T(0) g_{N})$ in Fig.~\ref{fig:glrat3} ($T(0)=\int dz \rho_{A}(b=0,z)$).
The deviation of this ratio from unity is an effect of nuclear shadowing at small $x$ and 
antishadowing for $0.03 < x < 0.2$.
The results of Fig.~\ref{fig:glrat3} should be compared with those of 
Fig.~\ref{fig:glrat}. By selecting DIS events with low impact parameters one 
can significantly increase nuclear shadowing. For instance, for $Q=2,5,10$ GeV and  $x=10^{-3}$ and $x=10^{-5}$, 
the increase is about $55 - 75 \%$ for $^{12}$C and $20 - 40 \%$ for $^{208}$Pb.

\subsection{The eikonal approximation approach}
\label{subsec:eik}

Another popular and frequently used approach to nuclear shadowing is the eikonal approximation. 
The essence of this method in DIS on nuclei is the assumption that the forward 
amplitude for the virtual photon interaction with a nucleus can be written as the probability 
of the transition $\gamma^{\ast} \to q \bar{q}$ (the virtual photon light-cone wavefunction) 
convoluted with an {\it exponential} factor describing the interaction of the  $q \bar{q}$ 
pair with the nucleons of the nucleus. The exponential factor is obtained by summing an 
infinite series of terms proportional to powers of the total $q \bar{q}$-nucleon scattering 
cross section (i.e., by {\it eikonalizing}, hence the name of the approximation).

An eikonalized form for high energy scattering amplitudes was first obtained by 
Cheng and Wu \cite{Wu}, for several processes in QED. 
In particular, these authors 
demonstrated that for processes such as Delbruck scattering (photon-nucleus scattering 
in the static approximation) the sum of Feynman graphs of a certain type can be cast in 
the form of a convolution of the $e^{+} e^{-}$-component of the virtual photon 
wavefunction with the eikonalized amplitude for the $e^{+} e^{-}$-nucleon scattering.

Two aspects of the QED derivation of \cite{Wu} are important to mention. 
Firstly, it was explicitly demonstrated that higher order diagrams, involving 
closed electron loops or graphs, in which  a photon is allowed to be emitted and 
absorbed by the same electron, do not exponentiate. Secondly, the transverse 
diameter of the $e^{+} e^{-}$-system emitted by the photon 
remains unchanged by the interaction. 

At ultra-relativistic energies, the eikonal approximation has been successful in 
the description of hadron-hadron scattering (for instance, the droplet model of 
\cite{Chou-Yang}) as well as hadron-nucleus scattering (see, for example, 
\cite{Glauber} and \cite{Gribovinel}). The approximation can be also applied to 
nucleus-nucleus scattering at high energies.

The use of the eikonal approximation in high energy DIS on nuclei,
within a dipole model for the wavefunction of a virtual photon,
constitutes an extension of its previous use in hadron-nucleus scattering and QED. 
As explained above, the forward virtual photon-nucleus amplitude is given by the square
of the photon light-cone wavefunction multiplied by an eikonal factor arising from the 
exponentiation of the {\it total} $q\bar{q}$-nucleon scattering cross section. 
Such a form is based on the strong assumption that the $q\bar{q}$-nucleon interaction 
leaves certain characteristics of the $q\bar{q}$-system, such as its transverse diameter and 
  momentum fraction sharing variable, unchanged, i.e., the 
interaction is diagonal in the appropriate variables. In other words, within the 
eikonal approximation the projectile is said to be frozen in its prepared state 
for the lifetime of the interaction.

A diagram typical of the eikonal approximation, as applied to nuclear shadowing, is shown in 
Fig.~\ref{fig:eik}. It represents the imaginary part of the forward virtual photon-nucleus
scattering amplitude, in which the interaction with only a pair of nucleons is taken into 
account. Such a graph gives rise to the shadowing correction, $\delta F_{2}^{A(2)}$, and should 
be compared to Fig.~\ref{fig:deu} of the leading twist approach. The differences between 
the two figures reveal conceptual differences between the approaches: while in 
Fig~\ref{fig:deu} the virtual photon interacts by dissociating into a multitude of 
diffractive intermediate states, the $q \bar{q}$-dipole of the virtual photon in 
Fig.~\ref{fig:eik} interacts elastically through the exchange of an interacting gluon ladder 
(illustrated by the exchange of two gluons).

\EPSFIGURE{eik2.epsi,height=6cm,width=9cm}{
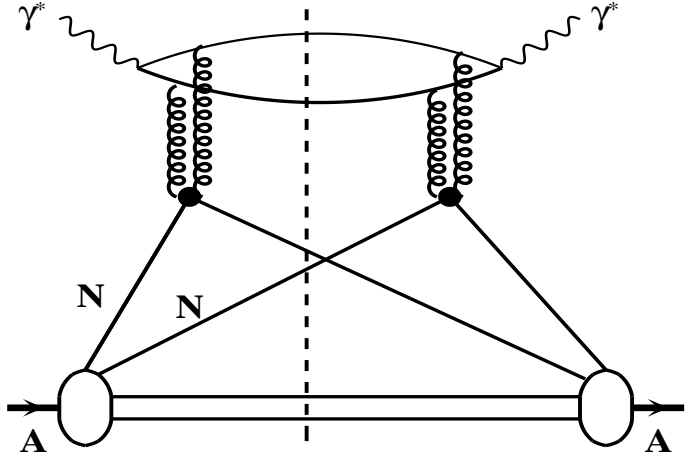
Feynman diagram giving rise to nuclear shadowing within the eikonal approximation.
\label{fig:eik}}

In quantum mechanical processes, and high-energy hadron-nucleus
scattering, the approximation that the dominant fluctuations of the
projectile can be considered to be frozen, can be justified by a suitable
choice of basis states which diagonalize the scattering operator. 
However, this is not the case in QCD, where we are already
given the basis, i.e., the quark and gluon degrees of freedom. In QCD, the
$q\bar{q}$-fluctuation of the virtual photon necessarily emits
gluons, which can emit further partons, etc. The Fock components of
the virtual photon, $|q \bar{q} \rangle$, $|q \bar{q} g \rangle$,
$\dots$, intermingle before and during the interaction with the target, 
i.e., those states are not eigenstates of the scattering operator and cannot in 
general be considered to be frozen. 
Intuitively, the observation that emitting extra gluons would lead to the breakdown of the 
eikonal approximation is similar to the finding of \cite{Wu} that the graphs 
where a photon is emitted and absorbed by the same fermion line, do not eikonalize.

Since the eikonal approximation is so successful in describing cross sections of 
nucleon-nucleus scattering, should it not also be a good method to describe 
lepton-nucleus DIS? The above discussion shows that the eikonal approximation 
is justified only for processes where the presence of the $|q \bar{q} g\rangle$-component 
of the virtual photon is unimportant. An example of a relevant observable 
is $F_{2}^{A}$ at $Q^2$ of only a few GeV$^2$. On the other hand, for reactions 
sensitive to the $|q \bar{q} g \rangle$-component, the eikonal approximation is 
expected to fail and give wrong size and $Q^2$-dependence of nuclear shadowing. 
One example of an observable for which this effect is dramatic is the 
longitudinal structure function $F_{L}^{A}$.

Within the eikonal approximation, the shadowing correction to the nuclear inclusive 
structure function $F_{2}^{A}$, $\delta F_{2}^{A}$, can be written in the form \cite{Bauer78}
\begin{eqnarray}
&&\delta F_{2}^{A}(x,Q^2) = \frac{Q^2}{4 \pi^2 \alpha_{em}} Re \Bigg[\int d\alpha ~d^2 d_{t} \sum_{i}~|\Psi(\alpha,Q^2,d_{t}^2,m_{i})|^2 ~\frac{A(A-1)}{2} \times \nonumber\\
&& ~\int d^2b \int_{-\infty}^{\infty} dz_1 \int_{z_{1}}^{\infty} dz_2 
(1-i\eta)^2 \Big[\sigma^{tot}_{q\bar q N}(x,d_{\perp}^2,m_{i})\Big]^2 ~\rho_A(b,z_1) ~\rho_A(b,z_2) e^{i2x m_N(z_1-z_2)} \times \nonumber \\
&&e^{-(A/2)(1-i\eta) \sigma^{tot}_{q\bar q N}(x,d_{\perp}^2,m_{i}) \int^{z_{2}}_{z_{1}} dz \rho_{A}(z)} \Bigg] \ .
\label{eik1}
\end{eqnarray}
Here $\alpha_{em}$ is the fine-structure constant; $\alpha$ is the fraction of the photon's 
longitudinal momentum carried by $q$ or $\bar{q}$; $d_{t}$ is the transverse diameter of 
the $q \bar{q}$-system; $m_{i}$ is the mass of the quark
with flavor $i$; $\rho_A$ is the nuclear one-body density. 
The square of the light-cone wavefunctions of the virtual photon is given by the standard expression
\begin{eqnarray}
|\Psi(\alpha,Q^2,d_{t}^2,m_{i})|^2&=&\frac{6~\alpha_{em}}{\pi^2} ~\sum_{i} e_{i}^2 \Bigg[\Big(Q^2 \alpha^2(1-\alpha)^2+ \frac{m_{i}^2}{4} \Big) K_{0}^2(\epsilon_{i} d_{t})  \nonumber\\
&+&\frac{1}{4}\Big(\alpha^2+(1-\alpha)^2\Big) ~\epsilon_{i}^2 ~K_{1}^2(\epsilon_{i} d_{t})\Bigg] \ ,
\label{wf}
\end{eqnarray}
where $K_{0}$ and $K_{1}$ are the modified Hankel functions; $\epsilon_{i}=Q^2\alpha(1-\alpha)+m_{i}^2$; $\eta$ is the ratio of the real to imaginary parts of $q \bar{q}$-nucleon 
scattering amplitude. In our analysis, we set $\eta=0.25$, which is consistent both 
with our leading twist calculations and vector meson dominance ideas, 
see e.g., \cite{Bauer78}. 
Note that since $F_{2}$ measures the virtual photon-target cross section averaged over helicities of the incoming photon, averaging over the photon helicities is assumed in Eqs.~(\ref{eik1}) and (\ref{wf}). Following our analysis in \cite{mfgs}, we include four flavors of quarks and assume  $m_{u}=m_{d}=m_{s}=300$ MeV and $m_{c}=1.5$ GeV.

The use of the optical theorem for the elementary $q \bar{q}$-nucleon amplitudes enables 
one to express the answer in Eq.~(\ref{eik1}) through the total $q \bar{q}$-nucleon 
cross section, $\sigma^{tot}_{q\bar q N}$. This cross section plays a key role in 
the so-called dipole formalism of DIS, and we refer the reader to \cite{amirim} for 
a brief review of various formulations of the dipole formalism existing in the literature.
In this analysis, we used a QCD-motivated  model for $\sigma^{tot}_{q\bar q N}$  developed in \cite{mfgs}. 
For small dipoles, 
the total cross section is predominantly inelastic and, as such,
is governed by the gluon distribution in the proton. 
In the non-perturbative region of large dipole sizes, we model 
$\sigma^{tot}_{q\bar q N}$ by requiring its equivalence to the soft pion-nucleon 
total cross section.

In exactly the same or very similar form, Eq.~(\ref{eik1}) was used in a number of 
papers \cite{FKS96,Nikolaev,Kop,KRT,Gotsman1,Gotsman2}. In the derivation of 
Eq.~(\ref{eik1}) one assumes that the invariant mass of all $q \bar{q}$-dipole 
intermediate states, which is inversely proportional to the dipole diameter $d_{t}$, is the same and approximately equal to  $Q$ 
(i.e., the diffractive variable $\beta = Q^2/(Q^2 + M^2) \approx 0.5$, which implies $x_{\Pomeron} \approx 2 x$). 
This means that the non-zero longitudinal momentum transfer to each individual nucleon 
of the nuclear target is of the order of $2 x m_{N}$, which explains the factor  
$\exp(i 2 xm_N (z_1 - z_2))$
 in Eq.~(\ref{eik1}).

We would like to stress the following subtle point, which should be kept in mind while 
applying Eq.~(\ref{eik1}) and which was ignored 
in \cite{Nikolaev, Kop, KRT, Gotsman1, Gotsman2}. The total cross section 
$\sigma^{tot}_{q\bar q N}$ receives contributions from both 
elastic, $\sigma^{el}_{q\bar q N}$, and inelastic, $\sigma^{inel}_{q\bar q N}$, 
cross sections. For small dipole sizes and not very small $x$, $\sigma^{inel}_{q\bar q N}$ 
is relatively small and can be calculated using methods of perturbative QCD. 
Moreover, in this case $\sigma^{el}_{q\bar q N}$ is negligibly small and thus 
one can identify $\sigma^{tot}_{q\bar q N}$ with $\sigma^{inel}_{q\bar q N}$. 
In circumstances when $\sigma^{inel}_{q\bar q N}$ and $\sigma^{el}_{q\bar q N}$ 
become sufficiently large, it is incorrect to ignore the contribution 
of $\sigma^{el}_{q\bar q N}$ to the total cross section. For example, 
$\sigma^{inel}_{q\bar q N}$ and $\sigma^{el}_{q\bar q N}$ become equal in the 
limit when the $q \bar{q}$-nucleon scattering amplitude reaches its limiting 
value allowed by unitarity of the scattering operator.

In order to illustrate how the eikonal approximation works, 
we apply Eq.~(\ref{eik1}) in the most straightforward way by assuming 
it is valid at all $Q^2$. Of course, as explained previously, 
this becomes progressively less justified as $Q^2$ increases because 
of the mixing of $|q \bar{q} \rangle$,  $|q \bar{q}g \rangle$ and 
$|q \bar{q}gg\ldots \rangle $ fluctuations of the virtual photon 
(in practice this restricts legitimate values of $Q^2$ to a few GeV$^2$).

\EPSFIGURE{f2aeiknew.epsi,height=13cm,width=15cm}{
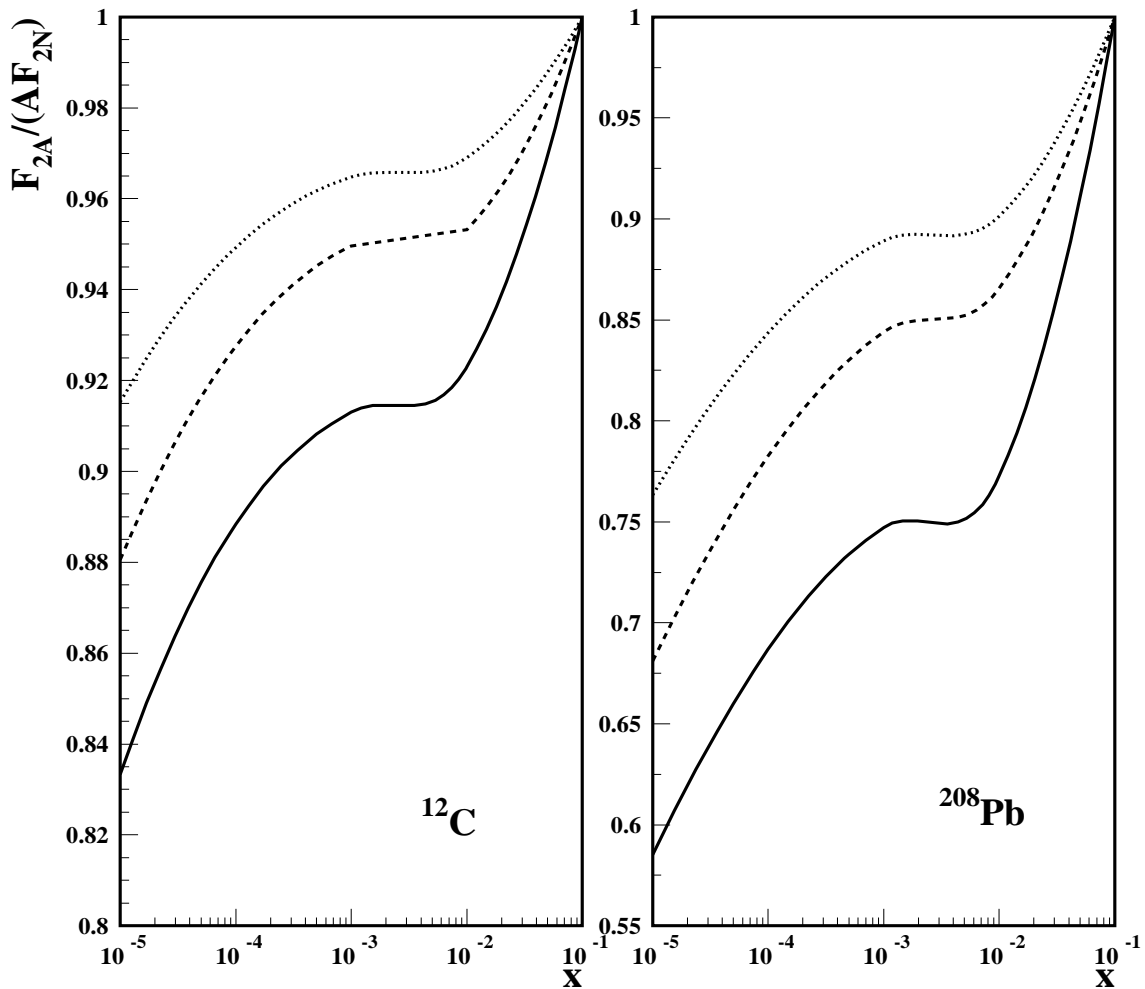
The ratio of nuclear to nucleon inclusive structure functions 
$F_{2}^{A}/(A F_{2}^{N})$ calculated within the eikonal approximation 
using Eq.~(\ref{eik1}). The $Q^2$-dependence is given by the solid ($Q=2$ GeV), 
dashed ($Q=5$ GeV) and dotted ($Q=10$ GeV) curves. \label{fig:eikf2}}

As it stands, Eq.~(\ref{eik1}) overestimates nuclear shadowing at the higher 
end of the shadowing region, $0.01 \leq x \leq 0.07$. In particular, 
$\delta F_{2}^{A}$ does not vanish at $x=0.1$, as happens in the leading 
twist approximation. Thus, in order to obtain sensible results with Eq.~(\ref{eik1}), 
it is applied only for $10^{-5} \leq x \leq 0.01$. For the interval 
$0.01 \leq x \leq 0.1$, we assumed that $\delta F_{2}^{A}$ decreases 
linearly and becomes zero at $x = 0.1$.

The ratio $F_{2}^{A}/(A F_{2}^{N})$, calculated using Eq.~(\ref{eik1}), 
is presented in Fig.~\ref{fig:eikf2} for $^{12}$C and $^{208}$Pb.
The $Q^2$-dependence of this ratio is shown by solid ($Q=2$ GeV), dashed 
($Q=5$ GeV) and dotted ($Q=10$ GeV) curves.
A comparison of Figs.~\ref{fig:eikf2} and \ref{fig:F2A} reveals the following 
characteristic trends (from the considered scenarios of nuclear shadowing within 
the leading twist approach, we choose the one 
corresponding to the H1 parameterization). 
At $Q = 2$~GeV, and for the $^{12}$C and  $^{208}$Pb nuclei, both leading twist and eikonal 
approaches give similar 
(within about 20\%)
 predictions for $F_{2}^{A}/(A F_{2}^{N})$. 
As $Q^2$ increases, for light and heavy nuclei the leading twist approximation predicts much larger shadowing than the eikonal approach. 
In particular, at $Q=10$ GeV and $x=10^{-3}$, $F_{2}^{A}/(A F_{2}^{N})$ is shadowed more in the leading twist approach than in the eikonal approximation by 
74\%
 for $^{12}$C and by 
71\%
 for $^{208}$Pb.
At $x=10^{-5}$, the corresponding increase of nuclear shadowing is similar for $^{12}$C and more modest for $^{208}$Pb: it is 
72\% for  $^{12}$C and 49\% for $^{208}$Pb.
The $Q^2$-behaviour follows the expected pattern: while the $Q^2$-behaviour of 
$F_{2}^{A}/(A F_{2}^{N})$ within the leading twist approach is governed by the 
QCD evolution equation, and is therefore logarithmic,  within the eikonal approximation it decreases with increasing $Q^2$ much faster and is dictated 
largely by the $Q^2$-dependence of the virtual photon light-cone wavefunction.

\subsection{Differences between the leading twist and eikonal approaches}
\label{subsec:com}

In this subsection, we summarize what has been said so far about the differences between 
the leading twist and eikonal approaches to nuclear shadowing in DIS on nuclei as well 
as discuss problems with an unambiguous implementation of the eikonal approximation.

The two key differences, which make the approaches so distinct, are related. They are: 
\begin{itemize}
\item{a different space-time evolution of the scattering process;}
\item{the neglect of the $|q \bar{q} g \rangle$ component (and higher Fock states) of 
the virtual photon in the eikonal approximation.}
\end{itemize}

The eikonal approach is reliable only within the framework of
non-relativistic quantum mechanics, where the number of interacting particles is conserved during collisions. In this case, the approximation that the Fock states of the incoming high-energy virtual photon can be considered frozen, is justified, and the procedure of eikonalization can be successfully carried out.

In contrast, in a quantum field theory such as QCD, the number of bare particles is not conserved.
In other words, the number of effective degrees of freedom, or relevant Fock states, 
in the photon wavefunction depends on $x$ and $Q^2$.
For example, the interacting $|q \bar{q} \rangle$ Fock state radiates gluons, 
thus creating and mixing with $|q \bar{q} g \rangle$, $|q \bar{q} g \dots g \rangle$ states.
This mixing is properly taken into account by the QCD evolution in the leading twist approximation. Of course QED is also a quantum field theory, however, the electromagnetic 
coupling is much weaker than the strong coupling of QCD, and there is no self-interaction 
of photons, so the corrections which spoil the eikonal approximation for 
QCD are correspondingly much smaller in the QED case.

One immediate consequence of the contrasting space-time evolution pictures within the
leading twist and eikonal approaches is the $Q^2$-dependence of nuclear shadowing. As
$Q^2$ increases, the Fock components of the virtual photon with an increasing number
of gluons, $|q \bar{q} g \dots g \rangle$, become important
for nuclear shadowing. This follows straightforwardly from the connection
between nuclear shadowing and gluon-dominated diffraction, as found in ZEUS
and H1 experiments at HERA. As a result, using the factorization theorem,
the $Q^2$-dependence of nuclear shadowing is governed by the DGLAP evolution
equation within the leading twist approach.

In the variant of the eikonal approximation that we considered, the
$|q \bar{q} g \rangle$ Fock state of the virtual photon is
absent. Hence, even if one adjusts the $|q \bar{q} \rangle$-nucleon
cross section in order to reproduce correctly the nuclear structure
function $F_{2}^{A}$ at the initial scale $Q_{0}$, the approach would
necessarily underestimate nuclear shadowing at large $Q^2$ since the
eikonal approximation is not based on QCD evolution.
This was already demonstrated by comparing Figs.~\ref{fig:eikf2} and \ref{fig:F2A}. 
As we will illustrate in the next section, this effect is even more dramatic for the longitudinal
structure function $F_{L}^{A}$.

The naive formulation of the eikonal model, which only contains
the $q \bar q$-component of the virtual photon wavefunction,
underestimates the amount of nuclear shadowing since it neglects diffractively
produced inelastic states, such as $q \bar q g$, $q \bar q gg$, etc.
The inclusion of a $q \bar q g$ component, as was done in the analysis
of diffraction on the nucleon in \cite{kgbdiff}, is only a step in the 
right direction. In order to reproduce the correct $Q^2$-behaviour of 
nuclear shadowing, which is governed by the DGLAP evolution equation, 
one should include the complete set of Fock states, 
i.e., an infinite series of components including infinitely many 
constituents.

There are several subtle points and technical problems with implementation 
of the eikonal approximation. Firstly, one has to be careful to use the 
total $q \bar{q}$-nucleon cross section in Eq.~(\ref{eik1}). In the kinematics, 
where the elastic and inelastic $q \bar{q}$-nucleon cross sections are 
compatible, the use of $\sigma_{q \bar{q} N}^{{\rm inel}}$ alone would significantly 
underestimate nuclear shadowing.

Secondly, in order to reproduce correctly nuclear shadowing at the higher end
of shadowing region, $0.01 \leq x \leq 0.07$, one needs to take into account 
the non-zero longitudinal momentum transfer to the nucleus through the factor 
$\exp (i 2 x m_{N} (z_{1}-z_{2}))$. In order to arrive at this factor in the eikonal 
approximation, one needs to make a strong assumption that all essential Fock 
states of the virtual photon have the same invariant mass of the order 
of $Q$.

Thirdly and very importantly, there is no unambiguous way to 
generalize Eq.~(\ref{eik1}) for nuclear structure functions to something similar
to Eq.~(\ref{deltaf}) for individual nuclear parton distributions. 
Of course, one can attempt to replace the ratio $F_{2}^{A}/(AF_{2}^{N})$ 
in Eq.~(\ref{eik1}) by $f_{j/A}/(A f_{j/N})$ but what can one use for 
cross section $\sigma^{tot}_{q\bar q N}$ ?
The eikonal approximation gives no clue as to $\sigma^{tot}_{q\bar q N}$ 
for different flavors of partons. Moreover, since such a picture would not 
be based on the factorization theorem and QCD evolution, the scaling violations 
of quark and gluon nuclear distributions would not be consistent 
with each other and the DGLAP equations.

To conclude we would like to give a clear answer to the question
as to why the eikonal approximation works so well for hadron-nucleus
processes but fails for lepton-nucleus DIS processes.
For high energies the hadronic projectile fluctuates into configurations 
which subsequently interact with the target with 
similar cross sections. This means that the distribution over
cross section fluctuations is rather narrow. 
As a result, in hadron-nucleon scattering inelastic diffraction is a
small correction to elastic scattering. When these ideas are applied to
hadron-nucleus scattering, one sees that inelastic intermediate states 
(corresponding to inelastic diffraction) give a small contribution 
compared to elastic intermediate states (corresponding to elastic scattering). 
Hence, the eikonal approximation works well.

In lepton-nucleus DIS the situation is very different. Firstly, both small- and
large-size fluctuations of the virtual photon contribute to DIS on the nucleon,
which means that cross section fluctuations are very significant. Secondly,
for small-size fluctuations, the situation for DIS off a nucleon is opposite
to that assumed in the eikonal approximation: in the former the cross section for inelastic 
diffraction (a leading twist observable) 
exceeds\footnote{The reason for the dominance of inelastic diffraction 
with an increase of $Q^2$ at fixed small $x$ can be traced 
to the definition of elastic cross section, which was introduced using  
$t$-channel factorization so that the virtual photon-target cross section
has the form of the convolution of the photon ($q\bar q $) wavefunction
with the $q\bar q $-target cross section.
However, as
gluons are attached to the $q\bar q$ pair at large $Q$, they are
predominantly emitted well before the photon wave 
packet has approached the target. These configurations containing gluons, 
which on average have a large transverse size and may be in the color octet 
dipole state rather than in the color triplet dipole state, 
much more readily rescatter diffractively (i.e.,  with a small momentum transfer to
the nucleon). As a result, the genuine total cross section of the
interaction, as measured through double rescattering, turns out
to be much larger at large $Q^2$ than the cross section defined
via the elastic $q\bar q N$ cross section.}
the cross section for elastic scattering (a sub-leading twist observable). 
For the latter the reverse is true. 

As a result, inelastic intermediate states dominate in lepton-nucleus 
DIS initiated by small dipoles. Of course, the presence of large-size dipoles 
in the virtual photon wavefunction introduces a certain degree of similarity 
between hadron-nucleus and lepton-nucleus scattering, but this is not sufficient 
to justify the application of the eikonal approximation to DIS on nuclei.

\section{Nuclear shadowing of the longitudinal structure function $F_{L}^{A}$}

\label{sec:longitudinal}

As discussed above, the crucial difference between the leading twist and eikonal approaches 
is contrasting space-time evolution pictures of the scattering process. 
As a consequence, the higher Fock components of the virtual photon, containing gluons, 
are effectively included in the leading twist approach and are neglected in the eikonal 
approximation. Therefore any observable which is sensitive to the 
gluon distribution in the nucleus should be a good tool to distinguish 
between the leading twist and eikonal approximations.
We have already demonstrated this using the inclusive structure function, $F_{2}^{A}$, 
at large $Q^2$. An even more striking example is given by the longitudinal nuclear structure 
function, $F_{L}^{A}$, which is measured by DIS of longitudinally-polarized 
photons on nuclei. 
Other relevant processes include exclusive electroproduction of $\rho$
(dominated by longitudinally-polarized photons) and $J/\psi$ mesons off nuclei.

The nuclear structure function $F_{L}^{A}$ can be obtained by a straightforward generalization of the one-loop perturbative QCD result for the nucleon \cite{QCDbook}:
\begin{equation}
F^{A}_{L}(x,Q^2)=\frac{2 \alpha_{s}(Q^2)}{\pi} \int^{1}_{x} \frac{dy}{y} (\frac{x}{y})^2 \Big(\sum^{n_{f}}_{i=1} e_{i}^2  (1-\frac{x}{y}) y g_{A}(y,Q^2)+\frac{2}{3}F_{2}^{A}(y,Q^2)\Big) \ .
\label{eq:fl_lt}
\end{equation}
Here $n_{f}$ is the number of active quark flavors. Replacing $g_{A}$ by $g_{N}$ and $F_{2}^{A}$ by $F_{2}^{N}$, one can present the ratio of the nuclear to nucleon longitudinal structure functions, $F^{A}_{L}/(A F_{L}^{N})$, in the form
\begin{equation}
\frac{F^{A}_{L}(x,Q^2)}{A F_{L}^{N}(x,Q^2)}=\frac{\int^{1}_{x} \frac{dy}{y} (\frac{x}{y})^2 \Big(\sum^{n_{f}}_{i=1} e_{i}^2  (1-\frac{x}{y}) y g_{A}(y,Q^2)+\frac{2}{3}F_{2}^{A}(y,Q^2)\Big)}{\int^{1}_{x} \frac{dy}{y} (\frac{x}{y})^2 \Big(\sum^{n_{f}}_{i=1} e_{i}^2  (1-\frac{x}{y}) y g_{N}(y,Q^2)+\frac{2}{3}F_{2}^{N}(y,Q^2)\Big)} \ .
\label{eq:fl_lt2}
\end{equation}

\EPSFIGURE{flratnew.epsi,height=13cm,width=15cm}{
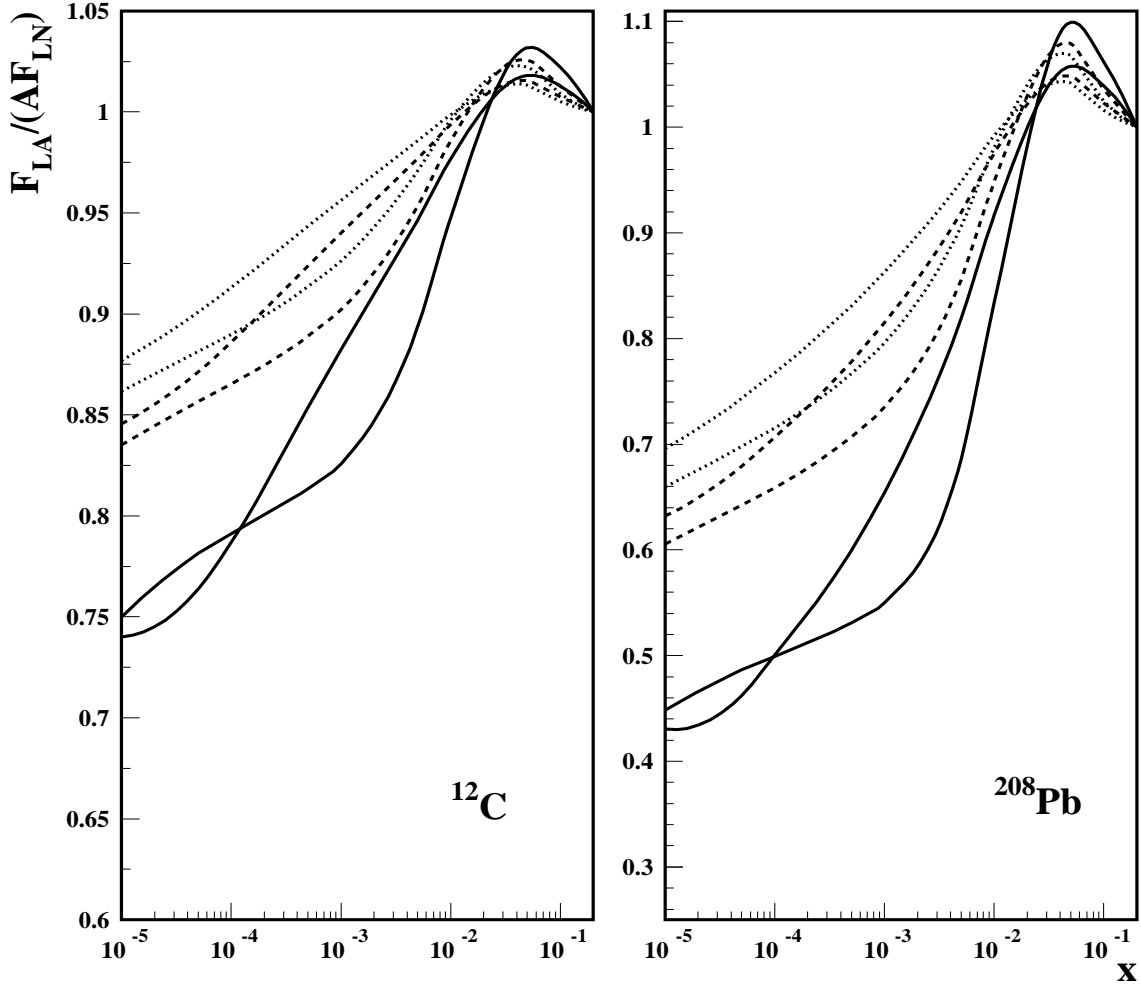
The ratio of nuclear to nucleon longitudinal structure functions $F_{2}^{L}/(A F_{2}^{L})$ calculated within the leading twist approximation using Eq.~(\ref{eq:fl_lt2}).
Different curves correspond to different values of $Q^2$ and scenarios of shadowing as in Fig.~\ref{fig:glrat}. \label{fig:FLA}}

The ratio $F_{L}^{A}/(A F_{L}^{N})$, calculated using Eq.~(\ref{eq:fl_lt2}), 
is presented in Fig.~\ref{fig:FLA} (curves are labelled as in Fig. \ref{fig:glrat}). 
By comparing Fig.~\ref{fig:FLA} to Fig.~\ref{fig:glrat} one sees that the size 
and $Q^2$-dependence of nuclear shadowing in the ratio $F_{L}^{A}/(A F_{L}^{N})$ 
are similar to those in the ratio $g_{A}/(A g_{N})$. Slightly smaller shadowing for 
$F_{L}^{A}/(A F_{L}^{N})$ is an effect of the convolution and sea quarks present 
in Eq.~(\ref{eq:fl_lt2}).

Within the eikonal approximation, one cannot use Eq.~(\ref{eq:fl_lt2}) because the eikonal approximation does not present an unambiguous scheme to evaluate $g_{A}$. Instead of Eq.~(\ref{eq:fl_lt2}),
the longitudinal structure function $F_{L}^{A}$ in the eikonal approximation can be obtained using Eq.~(\ref{eik1}) by replacing the helicity-averaged photon wavefunction $\Psi$ by the longitudinally-polarized one $\Psi_{L}$:
\begin{equation}
|\Psi_{L}(\alpha,Q^2,d_{t}^2,m_{i})|^2=\frac{6 \alpha_{em}}{\pi^2} \sum_{i} e_{i}^2 Q^2 \alpha^2(1-\alpha)^2 K_{0}^2(\epsilon_{i} d_{t}) \ .
\label{wfl}
\end{equation}
Then the ratio $F_{L}^{A}/(A F_{L}^{N})$ can be presented in the form
\begin{eqnarray}
&&\frac{F_{L}^{A}(x,Q^2)}{AF_{L}^{N}(x,Q^2)}=1-\frac{A-1}{2} Re \Bigg[
\int d\alpha d^2 d_{t} \sum_{i}|\Psi_{L}(\alpha,Q^2,d_{t}^2)|^2 \int d^2b \int_{-\infty}^{\infty} dz_1 \int_{z_{1}}^{\infty} dz_2 \times \nonumber \\
&&\Big(\sigma^{tot}_{q\bar q N}(x,d_{\perp}^2,m_{i})\Big)^2 ~\rho_A(b,z_1) ~\rho_A(b,z_2) e^{i 2 x m_N(z_1-z_2)} e^{-(A/2) (1-i\eta)\sigma^{tot}_{q\bar q N}(x,d_{\perp}^2,m_{i}) \int^{z_{2}}_{z_{1}} dz \rho_{A}(z)} \Bigg] \Big/ \nonumber \\
&&\Big(\int d\alpha d^2 d_{t}\sum_{i}|\Psi_{L}(\alpha,Q^2,d_{t}^2,m_{i})|^2 \sigma^{tot}_{q\bar q N}(x,d_{\perp}^2,m_{i})\Big) \ .
\label{eq:fl_eik}
\end{eqnarray}

Results for the ratio of the longitudinal structure functions in the eikonal approximation 
are given in Fig.~\ref{fig:fleik}. 
As for $F_{2}^{A}/(A F_{2}^{N})$ within the eikonal approximation, we assumed that
$F_{L}^{A}/(A F_{L}^{N})$ increases linearly for $x > 0.01$ and becomes unity at $x=0.1$.

\EPSFIGURE{flaeiknew.epsi,height=13cm,width=15cm}{
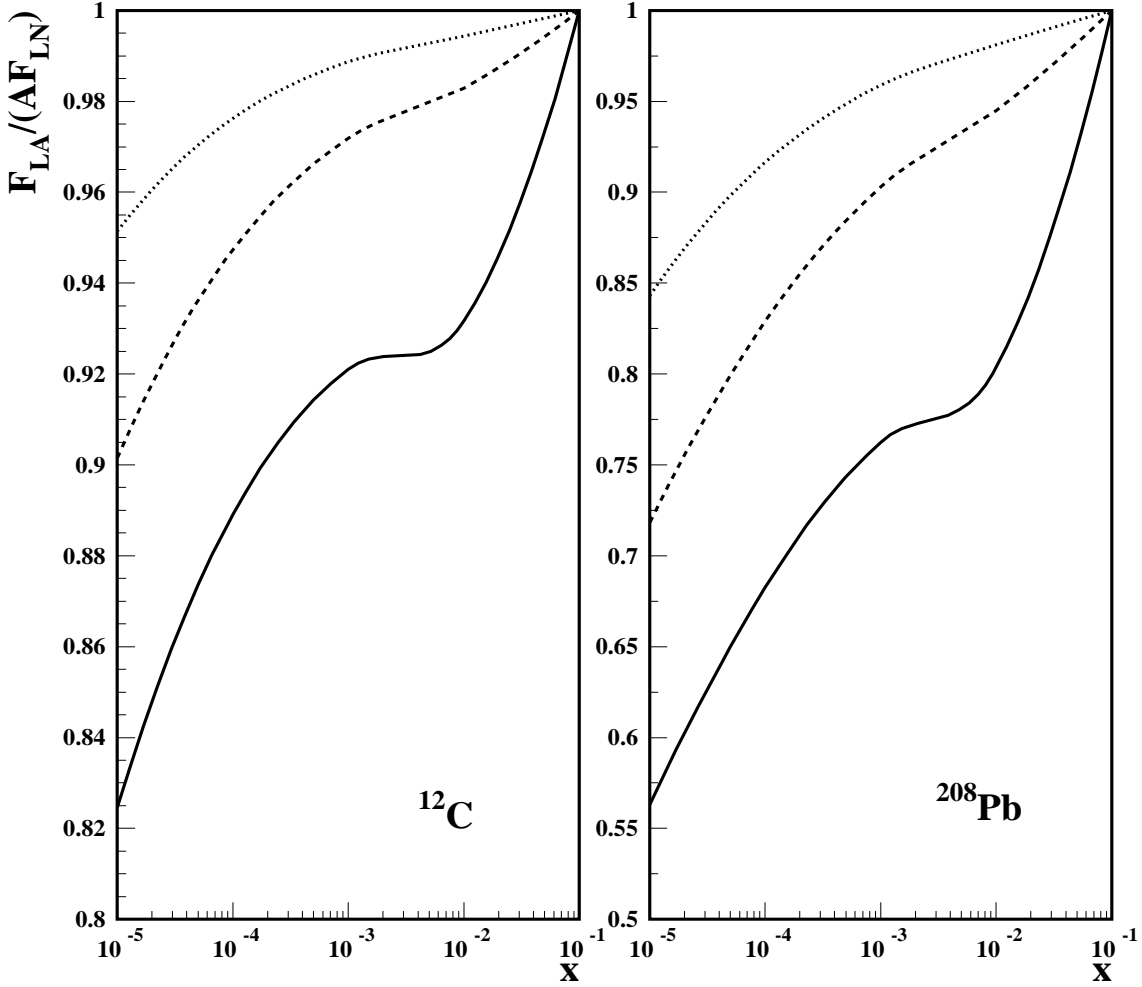
The ratio of nuclear to nucleon longitudinal structure functions $F_{2}^{L}/(A F_{2}^{L})$ calculated within the eikonal approximation using Eq.~(\ref{eq:fl_eik}).
The $Q^2$-dependence is shown by the solid ($Q$=2 GeV), dashed ($Q$=5 GeV) and dotted ($Q$=10 GeV) curves. \label{fig:fleik}}

The comparison of Figs.~\ref{fig:FLA} and \ref{fig:fleik} clearly reveals dramatic 
differences between the leading twist and eikonal approach predictions. 
Firstly, the leading twist approach already predicts much larger shadowing  
at the initial scale $Q_{0} = 2$~GeV. Secondly, nuclear shadowing dies out with 
increasing $Q^2$ much faster within the eikonal approximation. 
In order to quantify this discussion, our results for the nuclear shadowing 
correction to the longitudinal structure function, $1-F_{L}^{A}/(A F_{L}^{N})$, 
of $^{12}$C and $^{208}$Pb are presented in Table~1 at $x=10^{-3}$ and $x=10^{-5}$.
For the column containing the leading twist results, first values correspond to the ACWT parameterization (and the values in parenthesis to the H1 parameterization).
One can see from Table~\ref{tab:1} that, depending on $x$ and $A$, the shadowing correction to 
$F_{L}^{A}$ in the leading twist approach is larger, compared to the eikonal approximation, 
by a factor of $1.3 - 2$ at $Q = 2$~GeV and by a factor of two to seven at $Q=10$~GeV. 
Hence, measurements of $F_{L}^{A}$ should be an excellent means to distinguish between 
the leading twist and eikonal approaches 
to nuclear shadowing.

\TABULAR[ht]{|c|c|c|c|c|c|}{
\hline
 & $x$ & $Q$ (GeV) & $1-F_{L}^{A}/(A F_{L}^{N})$, leading twist & $1-F_{L}^{A}/(A F_{L}^{N})$, eikonal\\
\hline
 & & 2 & 0.25 (0.26) & 0.18 \\
 $^{12}$C & $10^{-5}$ &  5 & 0.16 (0.15) & 0.10 \\
 & & 10 & 0.14 (0.12) & 0.049 \\
\hline
& & 2 & 0.17 (0.12) & 0.079 \\
 $^{12}$C & $10^{-3}$ &  5 & 0.098 (0.060) & 0.028 \\
 & & 10 & 0.074 (0.044) & 0.011 \\
\hline
\hline
 & & 2 & 0.55 (0.57) & 0.43 \\
 $^{208}$Pb & $10^{-5}$   & 5 & 0.39 (0.37) & 0.28 \\
 & & 10 & 0.34 (0.30) & 0.16 \\
\hline
& & 2 & 0.45 (0.35) & 0.24 \\
 $^{208}$Pb & $10^{-3}$   & 5 & 0.26 (0.18) & 0.097 \\
 & & 10 & 0.20 (0.14) & 0.041 \\
\hline
}{Nuclear shadowing correction to the longitudinal structure function within the leading twist and eikonal approaches. \label{tab:1}}

\section{Conclusions and discussion}
\label{sec:conc}

We compare two frequently used approaches to nuclear shadowing in DIS from nuclei: 
the leading twist and eikonal approaches. Our comparison is based on the observation 
that one of the foundation principles of the two approaches, the space-time 
picture of the interaction, is quite different. The leading twist approach is based 
on the observation that at high energies the projectile interacts simultaneously with 
several nucleons of the nuclear target. This leads to the reference frame independent 
connection between nuclear shadowing in DIS from nuclei and DIS diffraction from the nucleon. 
The QCD factorization theorems for inclusive and diffractive scattering enable one to express 
the shadowing correction to the individual parton distribution functions in nuclei in terms 
of the corresponding diffractive parton distribution functions of the proton. 
Moreover, another consequence of the factorization theorems is that the scaling 
violations of nuclear parton distribution functions are given by the leading twist DGLAP equation. 
 
Unlike the leading twist approach, the eikonal approximation is a frame-dependent method, 
which applies in the target rest frame. In this reference frame, the incoming virtual 
photon first fluctuates into partonic components 
($|q \bar{q} \rangle$, $|q \bar{q}g \rangle$, $\dots$), which then interact with the nucleus. 
We considered the often-used approximation in which only the $|q \bar{q} \rangle$ fluctuation 
is considered. In this case, the $q \bar{q}$ effective dipole interacts successively 
and elastically with nucleons of the target. Apart from this, an additional assumption 
is made that such interactions can be presented in an exponential or eikonalized form. 
It is important to realise that the eikonal approximation implies that only elastic 
($|q \bar{q} \rangle$) intermediate states contribute to the virtual photon-nucleus 
cross section. In addition, the method is not based on the QCD factorization theorem and  
therefore does not include the proper QCD evolution. 

In this work, we performed several numerical studies.

Firstly, within the leading twist approximation, the ratio of the nuclear to 
nucleon gluon and sea quark distribution functions was examined as a function 
of Bjorken $x$ for several values of $Q$ and for a wide range of nuclei. 
By testing four distinct parameterizations of diffractive partons distributions 
of the proton, we extended the earlier analysis of \cite{FS99} and confirmed 
that the gluons in nuclei are shadowed more than the sea quarks. 
We found that the difference between gluon and sea quark 
nuclear shadowing was not quite as large as that found in \cite{FS99}
(compare Fig.~\ref{fig:F2A} to Fig.~5 of \cite{FS99} which uses the ACWT fit).
There are two major reasons for this. Firstly the H1-type fit corresponds  
to a smaller diffractive gluon density at large $\beta $ than ACWT. Secondly, 
in the current analysis we allowed for a possible difference of the slopes for 
the diffractive gluon and quark distributions (see appendix~\ref{appendix:a} for details).
In addition, we made predictions for shadowing of the gluon parton distributions 
at central impact parameters as well as for the charm parton distributions.

Secondly, we compared predictions for the nuclear structure functions, 
$F_{2}^{A}$, within the leading twist and eikonal approximations. 
One should emphasize two aspects of this comparison: 
the size and $Q^2$-dependence of the nuclear shadowing correction. 
At the initial scale $Q=2$ GeV, both leading twist and eikonal approaches 
give similar predictions. Nevertheless, the $Q^2$-dependence of $F_{2}^{A}$ gives 
good discriminating power between the leading twist and eikonal approaches 
which reflects their conceptual differences:  
a logarithmic decrease of nuclear shadowing in the 
leading twist approach is much slower than that predicted using the eikonal approximation. 
For instance, at $Q=10$ GeV and $x=10^{-3}$, the leading twist approach predicts 
71\% more shadowing for $F_{2}^{^{208}{\rm Pb}}$ than the eikonal approximation.

Thirdly, given the fact that the eikonal approximation omits contributions of the 
virtual photon fluctuations containing gluons ($|q \bar{q} g \rangle$, $\dots$) it must 
predict very small nuclear shadowing for observables sensitive to the 
gluon distributions of nuclei. 
Examples of such observables include cross sections for
longitudinally polarized $\rho$ meson electroproduction, 
$J/\psi$ electroproduction, and the longitudinal structure function $F_{L}^{A}$. 
We considered the latter in detail within the leading twist and eikonal approximations. 
The differences between the predictions for the size and $Q^2$-dependence of the shadowing 
correction to $F_{L}^{A}$ are very dramatic: the eikonal approximation undershoots the 
leading twist prediction for nuclear shadowing by at least a factor of two for all of the 
nuclei and values of $Q^2$ which we considered. Therefore, since $F_{L}^{A}$ can be 
measured at the future Electron-Ion Collider \cite{EIC}, this observable should be 
able to discriminate unambiguously between the leading twist and eikonal approaches 
to nuclear shadowing in DIS from nuclei.  

Nuclear shadowing is closely related to another interesting
small-$x$ phenomenon namely the violation of the DGLAP approximation at
sufficiently small values of $x$. 
This phenomenon is referred to by various names in the literature: 
parton saturation, parton taming, unitarity constraints, etc. 
Let us consider DIS from a heavy nucleus at very large energies, 
i.e., at very small $x$. For almost all essential impact parameters, 
the nuclear structure function is predicted to increase mildly with 
decreasing $x$, $F_{2}^{A} \propto \ln (1/x)$, which is significantly slower than 
the behaviour predicted by the DGLAP equation (see e.g., \cite{eathera}).
It is now understood that parton saturation and related phenomena occur in heavy nuclei 
at larger $x$ than in the proton, which justifies the use of nuclear beams in attempt 
to study new aspects of small-$x$ physics. On the other hand, the leading twist nuclear 
shadowing significantly reduces the density of partons (especially gluons) at 
small $x$ and thus competes with the all-twist phenomenon of parton saturation. 
As a result, sensible studies of QCD at high parton densities can be carried out 
only if nuclear shadowing of partons is properly taken into account.

\section*{\acknowlname}

The authors are  indebted to GIF, ARC, PPARC, and DOE for support. M.S.  thanks J.~Collins and F.~Hautmann for useful discussions.

% APPENDIX A

\begin{appendix}

\section{Parameterizations of diffractive parton
distributions and $\sigma_{eff}$}
\label{appendix:a}

Within the leading twist approach, the nuclear shadowing correction, $\delta f_{j/A}$, 
to nuclear PDFs is given by Eq.~(\ref{qe}). It involves the effective cross section, 
$\sigma^j_{eff}$, which gives the strength of the interaction with any two 
nucleons of the nucleus leading to nuclear shadowing of the parton of flavor $j$. 
This may be expressed through the nucleon diffractive parton distribution function (DPDF) for a parton of the same flavour as follows 
\cite{FS99,AFS99}:
\begin{eqnarray}
&&\sigma_{eff}^{j}(x,Q^2) = \frac{16 \pi}{f_{j/N}(x,Q^2)(1+\eta^2)} 
\int_x^{x_{\Pomeron, 0}} dx_{\Pomeron} f_{j/N}^{D(4)}(\beta, Q^2,x_{\Pomeron},t=0) \ ,
\label{sflt}
\end{eqnarray}
where $f_{j/N}$ is the 
inclusive parton distribution of the nucleon; $f_{j/N}^{D(4)}$ is the diffractive parton 
distribution of the nucleon (which is, strictly speaking, a proton);
$\eta$ is the ratio of the real to imaginary parts of the diffractive scattering amplitudes. In our analysis, $\eta=0.22$ for the amplitude corresponding to the ACWT and model 3 parameterizations and $\eta=0.32$ when the H1 parameterization is used.

As explained in Sect.~\ref{sec:main}, the upper limit of integration, 
$x_{\Pomeron, 0}$, is different for the gluons ($x_{\Pomeron, 0}=0.03$: to allow 
antishadowing) and for the sea quarks ($x_{\Pomeron, 0}=0.1$). However, for the sea quarks, 
because of the factor $\exp(i x_{\Pomeron}m_N(z_1-z_2))$ 
in Eq.~(\ref{qe}) and the absence of antishadowing, 
the exact value of $x_{\Pomeron, 0}$ is unimportant for our numerical analysis.

Ignoring the small minimum momentum transfer, $t_{min}$, always present in inelastic 
diffraction, the condition $t \approx -k^2_{t} = 0$ in Eq.~(\ref{sflt}) means that 
$f_{j/N}^{D(4)}$ is to be evaluated at $t=0$. However, all the parameterizations to 
the diffractive data which we used are fits to 
$t$-integrated diffractive structure function $F_{2}^{D(3)}$. 
This means that we should assume a certain $t$-dependence of $f_{j/N}^{D(4)}$. 
For practical reasons, we assumed a simple exponential dependence
\begin{equation}
f^{D(4)}_{j/N}(\beta, Q^2,x_{\Pomeron},t)=e^{-B(x_{\Pomeron}) |t|} f^{D(4)}_{j/N}(\beta, Q^2,x_{\Pomeron},t=0) \ ,
\end{equation}
where the slope $B$ could depend on $x_{\Pomeron}$ (for the sea quarks) as well as on 
Bjorken $x$ (for the gluons).

Introducing the diffractive distribution function $f^{D(3)}_{j/N}$ as
\begin{equation}
f^{D(3)}(\beta, Q^2,x_{\Pomeron}) \equiv \int^{0}_{-\infty} dt f^{D(4)}_{j/N}(\beta, Q^2,x_{\Pomeron},t) \ ,
\end{equation}
one readily obtains that 
\begin{equation}
f^{D(4)}_{j/N}(\beta, Q^2,x_{\Pomeron},t=0)=B(x_{\Pomeron}) f^{D(3)}_{j/N}(\beta, Q^2,x_{\Pomeron}) \ .
\label{d4d3}
\end{equation}

As a next step, we use the Regge factorization hypothesis 
which assumes that 
$f^{D(3)}_{j/N}$ is a product of a Pomeron flux factor, 
$f_{\Pomeron}(x_{\Pomeron})$, and the parton distribution functions of the Pomeron, 
$f_{j/\Pomeron}(\beta,Q^2)$:
\begin{equation}
f^{D(3)}_{j/N}(\beta, Q^2,x_{\Pomeron}) = 
f_{\Pomeron}(x_{\Pomeron})f_{j/\Pomeron}(\beta,Q^2) \ .
\label{regge}
\end{equation}
The comparison of Eqs.~(\ref{d4d3}) and (\ref{regge}) gives
\begin{equation}
f^{D}_{j/N}(\beta,Q^2,x_{\Pomeron},t=0) = B(x_{\Pomeron}) ~f_{\Pomeron}(x_{\Pomeron}) 
~f_{j/\Pomeron}(\beta,Q^2) \ .
\label{t0}
\end{equation}
Substituting Eq.~(\ref{t0}) into Eq.~(\ref{sflt}), we obtain our master equation for $\sigma_{eff}^{j}$: 
\begin{equation}
\sigma_{{\rm eff}}^{j}(x,Q^2) = \frac{16 \pi}{f_{j/N}(x,Q^2)(1+\eta^2)} 
\int_x^{x_{\Pomeron, 0}} dx_{\Pomeron} B(x_{\Pomeron}) ~f_{\Pomeron}(x_{\Pomeron}) ~f_{j/ \Pomeron}(\beta, Q^2) \ .
\label{sfltb}
\end{equation}

\EPSFIGURE{su_finalnew2w.epsi,height=12cm,width=12cm}{
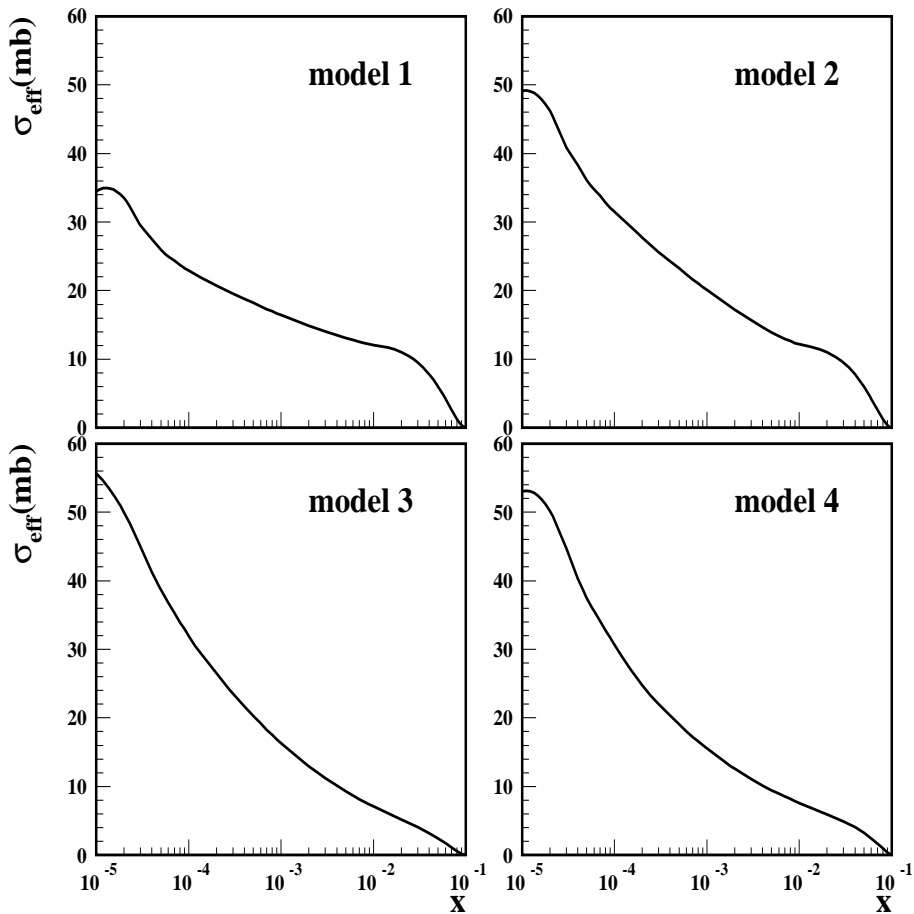
The effective cross section, $\sigma_{{\rm eff}}^{u}$, for the up sea quarks obtained 
using Eq.~(\ref{sfltb}) at $Q=2$ GeV. \label{fig:sigusea}}

\EPSFIGURE{sgl_finalnew2.epsi,height=12cm,width=12cm}{
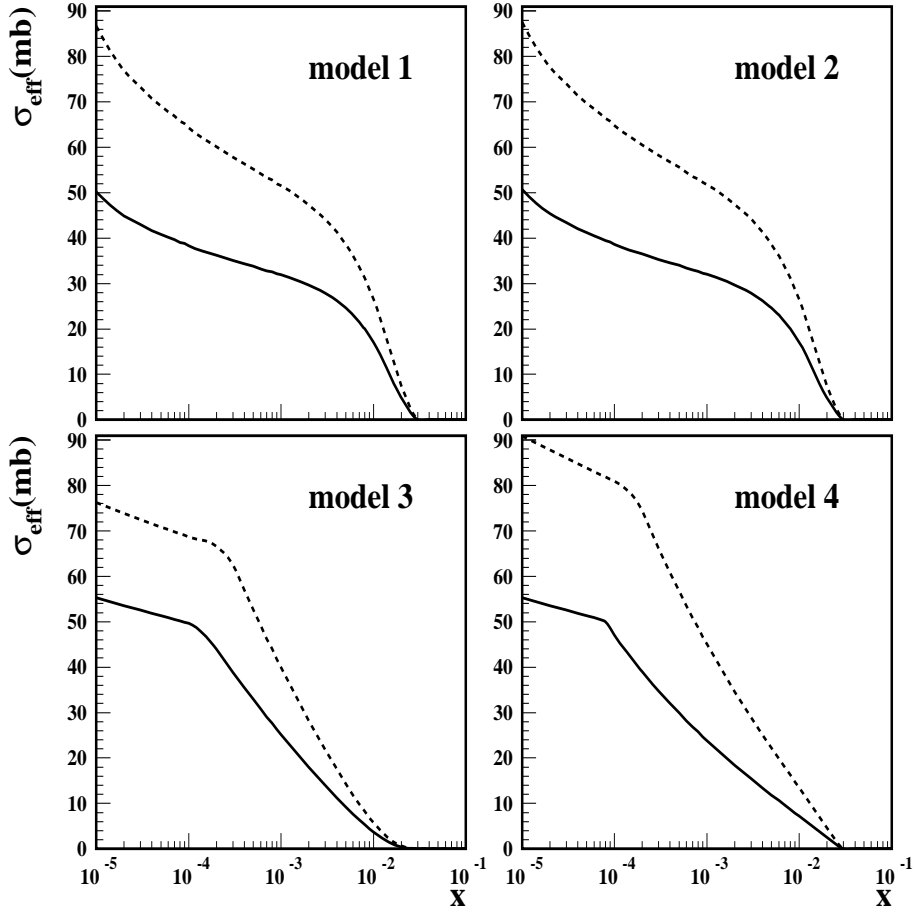
The effective cross section $\sigma_{{\rm eff}}^{g}$ for the gluons obtained using 
Eq.~(\ref{sfltb}) (with the unitarity restriction) at $Q=2$ GeV. The solid curves 
correspond to the slope $B_{g}$ given by Eq.~(\ref{eq:bg}), while the dashed curves 
are calculated using $B_{g}$ of Eq.~(\ref{slopeg2}). \label{fig:siggl} 
}

The theoretical analysis of the HERA diffractive data, in terms of quark and gluon degrees of freedom, with the additional assumption of Regge factorization and Pomeron dominance, effectively
concerns itself with the parton distributions in the Pomeron, $f_{j/\Pomeron}(\beta,Q^2)$. 
One way to study $f_{j/\Pomeron}$ phenomenologically is to fit the experimental data by 
performing QCD evolution using a reasonable trial shape for $f_{j/\Pomeron}$ at the 
initial scale $Q_{0}$. Models 1, 2 and 4, discussed below, are examples of such a 
determination of $f_{j/\Pomeron}$. Model 3 is a theoretical prediction for $f_{j/\Pomeron}$, 
which also successfully describes certain diffractive data. 

It is important to emphasize that the four considered parameterizations of the 
gluon distribution function in the Pomeron, $f_{g/\Pomeron}$, become so large at small 
values of Bjorken $x$ ($x < 3 \times 10^{-4}$) that the gluon-induced diffractive 
DIS cross section exceeds half the gluon-induced inclusive DIS cross section. 
Clearly, this is prohibited by the unitarity of the $S$ matrix (when the unitarity 
limit is reached, the diffractive and elastic cross sections are equal each other 
and are equal to half the total cross section). In terms of $f_{g/\Pomeron}$, the 
unitarity restriction can be written in the form
\begin{equation}
\int^{x_{\Pomeron, 0}}_{x} d x_{\Pomeron}f_{\Pomeron}(x_{\Pomeron}) 
f_{g/\Pomeron} (x/ x_{\Pomeron}, Q^2) \leq \frac{1}{2} g(x,Q^2) \ ,
\label{eq:unit}
\end{equation}
where $g(x,Q^2)$ is 
the number density of gluons in the nucleon.
So, whenever $f_{g/\Pomeron}$ becomes so large as to violate the unitarity 
limit of Eq.~(\ref{eq:unit}), $f_{g/\Pomeron}$ is replaced by its limiting value given 
by Eq.~(\ref{sfltb}). 
This allows $\sigma^g_{eff}$ to continue to grow as $x$ decreases but 
only at a reduced rate dictated by the growth of the inclusive gluon PDF.

Predictions for $\sigma_{{\rm eff}}$ for the up-quark sea and gluon, obtained using 
Eq.~(\ref{sfltb}) (with the unitarity restrictions on $\sigma_{{\rm eff}}^{g}$), are 
presented in Figs.~\ref{fig:sigusea} and \ref{fig:siggl}. 
All curves correspond to $Q=2$ GeV. For the inclusive parton distributions, the CTEQ5M parameterization \cite{CTEQ5} was used.
 For the gluons, 
we tested two choices of the diffractive slope $B_{g}$: the solid curves correspond 
to $B_{g}$ parametrised by Eq.~(\ref{eq:bg}), while the dotted curves 
are obtained with $B_{g}$ of Eq.~(\ref{slopeg2}) below.

We now describe the parameterizations of DPDFs, 
which we used, in detail.  

\medskip
\noindent {\bf Model 1}
\medskip

Numerical predictions in \cite{FS99} are based on the parameterization by Alvero, Collins, Terron and Whitmore \cite{ACWT}.
The choice of fit D gives the best fit to the data on diffractive DIS and diffractive
photoproduction of jets taken at HERA by the ZEUS and H1 collaborations.
Note that this parameterization somewhat overestimates the most recent
H1 data on jet production \cite{H1dijet}.

The following shapes of the quark ($q=u=\bar{u}=d=\bar{d}$) and gluon parton distribution 
functions of the Pomeron were obtained at the initial scale $Q_0 = 2$~GeV \cite{ACWT}
\begin{eqnarray}
&&\beta f_{q/\Pomeron}(\beta,Q_{0}^2)=0.292\Big(\beta(1-\beta)-0.159(1-\beta)^2\Big) \ , \nonumber\\
&&\beta f_{g/\Pomeron}(\beta,Q_{0}^2)=9.7\beta(1-\beta) \ .
\label{acwt1}
\end{eqnarray}
The strange and charm quark distributions are taken to be zero at the initial scale $Q_0$ and 
are generated by QCD evolution for $Q^2 > Q_{0}^2$.

We would like to point out two important features of the parameterization given by Eq.~(\ref{acwt1}).
Firstly, a successful DGLAP fit to the diffractive data requires that
the gluon distribution is much larger than the quark distribution, in order to reproduce 
the observed scaling behaviour in $Q^2$.
This automatically implies that $\sigma_{eff}^{g}$ is much larger than $\sigma_{eff}^{u}$ 
(e.g., compare Fig.~\ref{fig:sigusea} to Fig.~\ref{fig:siggl} in the region around $x=10^{-3}$). 
Secondly, the quark distribution is rather hard and it even becomes negative at low values of $\beta$. 
However, this does not lead to a paradox since the diffractive data with low $\beta$ were 
not used in the fitting procedure of \cite{ACWT}. 
Thus, the low-$\beta$ behaviour of the parameterization~(\ref{acwt1}) is not well constrained.

The Pomeron flux used in Eq.~(\ref{sfltb}) was parameterized in the Donnachie-Landshoff form
\begin{equation}
f_{\Pomeron}(x_{\Pomeron})=\frac{9 \beta_{0}^2}{4 \pi^2} \int^{0}_{-1} dt \Big[
\frac{4 m_{p}^2-2.8 t}{4 m_{p}^2-t}\big(\frac{1}{1-t/t_1}\big)^2\Big]^2
x_{\Pomeron}^{1-2 \alpha_{\Pomeron}(t)} \ ,
\label{dl}
\end{equation}
where $m_{p}$ is the proton mass; $\beta_{0}=1.8$ GeV$^{-1}$ is the Pomeron-quark coupling; 
$t_1=0.7$ GeV$^{2}$;
$\alpha_{\Pomeron}(t)=\alpha_{\Pomeron}+\alpha^{\prime}t$ is the Pomeron trajectory. 
The analysis of \cite{ACWT} showed that a successful fit the diffractive data 
favors $\alpha_{\Pomeron}=1.14$ and $\alpha^{\prime}=0.25$ GeV$^{-2}$.

The slope $B_{q}(x_{\Pomeron})$ for the quarks, which enters Eq.~(\ref{sfltb}), 
was found from the $t$-dependence of the Pomeron flux (see Eq.~(\ref{dl}))
\begin{equation}
B_{q}(x_{\Pomeron})=x_{\Pomeron}^{1-2 \alpha_{\Pomeron}(t=0)} \Big/ \int^{0}_{-1} 
dt \Big[\frac{4 m_{p}^2-2.8 t}{4 m_{p}^2-t} \big(\frac{1}{1-t/t_1}\big)^2\Big]^2
x_{\Pomeron}^{1-2 \alpha_{\Pomeron}(t)} \ ,
\label{slope}
\end{equation}
Thus defined $B_{q}(x_{\Pomeron})$ is a slow function of $x_{\Pomeron}$: $B_{q}$ increases 
with decreasing $x_{\Pomeron}$. For a wide range of $x_{\Pomeron}$, the value of $B_{q}$ 
is close to the slope of the diffractive structure function $F_{2}^{D(4)}$, 
$B = 7.2 \pm 1.1$~GeV$^{-2}$, which was measured by the ZEUS collaboration \cite{ZEUS2}.

For the gluon diffractive distribution, the slope $B_{g}$ is expected to be smaller than $B_{q}$.
The rationale for the use of a smaller slope for gluons in this and other models comes from 
two different lines of argument
\cite{fmsJPsi}
and \cite{hs}. 
In \cite{fmsJPsi} it was argued that $B_{g} \sim 5$~GeV$^{-2}$ leads to effective cross sections 
at $Q_0 \sim 2$~GeV similar to that which would be obtained by rescaling the quark-antiquark 
dipole cross sections we found in \cite{mfgs} by the Casimir operator factor of 9/4. 
Also we pointed out that $B_{g} \sim 7$~GeV$^{-2}$ would lead to $\sigma_{eff}$ for gluons as large 
as 60 mb, which would imply blackness for the interaction up to very large impact parameters. 
In \cite{hs}, based on the calculation of the diffractive parton densities, it was argued that 
the size of the dipole generating the gluon diffractive parton density should be small. This would 
also suggest a slope closer to the case of elastic $J/\psi$ photoproduction rather than to one 
expected if the soft physics dominates. 
Hence, in our analysis we used the 
$x$-dependent parameterization for $B_{g}$
given by Eq.~(\ref{eq:bg}).

In addition, in order to access the sensitivity of our results to the uncertainty in the 
choice of $B_{g}$, we considered the second option, when the diffractive slopes for the gluons 
and quarks are the same
\begin{equation}
B_{g}=B_{q} \ ,
\label{slopeg2}
\end{equation}
where $B_{q}$ is given by Eq.~(\ref{slope}).

\medskip
\noindent {\bf Model 2}
\medskip

We modified the parameterizations of Eq.~(\ref{acwt1}) by adding a low-$\beta$
piece to the Pomeron parton distributions,
to remove the rather unnatural feature that they go negative for small $\beta$.
As a guide, we used the factorization theorem for diffractive processes.
Since small values of $\beta$ correspond to large diffractive masses, $M^2 \gg Q^2 $,
the diffractive cross section is expected to be described by formulae derived in the 
triple Regge limit \cite{Mueller70} within which $W^2 \gg M^2 \gg Q^2$.
One assumes in this approach that multipomeron exchanges can be neglected.
Although there are no strong theoretical reasons to support this assumption  
the triple Regge formulae with an effective Pomeron exchange seem to work
pretty well up to rather large energies relevant for shadowing at $x \ge 10^{-4}$.
In the triple Regge approximation the ratio of the diffractive 
cross section at $t=0$ to total inclusive cross section in DIS 
should be the same as the ratio of diffractive and total cross section in real photon 
or hadronic diffraction. This may be expressed approximately 
(neglecting small corrections due to small deviations of $\alpha_{\Pomeron}$ from 1) as
\begin{equation}
\frac{d^4 \sigma^{diff}/dx dQ^2 dt dM_{X}^2 |_{t=0}}{d^2 \sigma /dx dQ^2}= \frac{A}{M_{X}^2} \ ,
\label{tail1}
\end{equation}
where the coefficient $A$ is process-independent, with a good accuracy. 
Indeed, data on real photon and pion diffractive dissociation on hydrogen
give $A = 0.122 \pm 0.006$~GeV$^{-2}$ for the photon and $A=0.118 \pm 0.006$ GeV$^{-2}$ 
for the pion \cite{Chapin}. An earlier experiment on pion, kaon and proton diffractive dissociation 
on hydrogen produced similar values of $A$: $A=0.102 \pm 0.013$~GeV$^{-2}$ for pions at 
$p_{lab} =200$~GeV/c, $A=0.094 \pm 0.005$~GeV$^{-2}$ for protons at $p_{lab} = 200$~GeV/c, 
and $A=0.097 \pm 0.027$~GeV$^{-2}$ for kaons at $p_{lab} = 100$~GeV/c \cite{Cool}. 
In our analysis, we use $A=0.12$~GeV$^{-2}$.

Assuming this low-$\beta$ tail appears once the quark distributions become zero, 
using Eq.~(\ref{tail1}) we arrive at the following for $\beta < \beta_0 = 0.137$:
\begin{equation}
\beta f_{q/\Pomeron}(\beta,Q_{0}^2) = \frac{0.02}{1-\beta} ~\frac{\beta_0 - \beta}{\beta_0}
\end{equation}
\noindent and use Eq.~(\ref{acwt1}) for larger $\beta > \beta_0$.
For the gluon channel the soft contribution is a small correction
hence we ignore it and implement Eq.~(\ref{acwt1}) for $\beta$.

\medskip
\noindent {\bf Model 3}
\medskip

The light-cone QCD model of Hautmann, Kunszt, and Soper \cite{HKS} leads to a set of the
Pomeron parton densities which, upon QCD evolution in $Q^2$, give a good
description of the ZEUS data. The quark ($q=u=\bar{u}=d=\bar{d}=s=\bar{s}$) and gluon 
parton distributions in the Pomeron, at the initial scale $Q_0 = 1.5$~GeV, read
\begin{eqnarray}
&&\beta f_{q/\Pomeron}(\beta,Q_{0}^2) = 0.0278 \beta ~\Big(1 + 0.824 \beta - 0.286 \beta^2 
- 2.713 \beta^3 + 1.218 \beta^4 \Big) \ , \nonumber\\
&&\beta f_{g/\Pomeron}(\beta,Q_{0}^2) = 0.987 ~\Big(1 + 0.821 \beta - 1.495 \beta^2 + 1.569 \beta^3 - 0.239 \beta^4 \Big) \ .
\label{hau1}
\end{eqnarray}
One can see from Eq.\ (\ref{hau1}) that, in contrast to the parameterizations of Eq.~(\ref{acwt1}), the low-$\beta$ region is 
treated in a sensible fashion.
In this model, the Pomeron flux, $f_{\Pomeron}$, is given by the following expression
\begin{equation} 
f_{\Pomeron}(x_{\Pomeron})=\frac{1}{x_{\Pomeron}}\Big(\frac{0.0042}{x_{\Pomeron}}\Big)^{0.253} \ .
\end{equation}
Note also that since this Pomeron flux is different from the one used in \cite{ACWT}, 
a direct comparison of the overall numerical coefficients in Eq.~(\ref{acwt1}) and Eq.~(\ref{hau1}) 
is not possible.

\medskip
\noindent {\bf Model 4}
\medskip

The H1 collaboration performed the QCD analysis of their own data \cite{H1} and produced 
an independent set of $f_{j/\Pomeron}$. The computer code with this parameterization of 
$f_{j/\Pomeron}$ is available from \cite{H1an}.
One should note that this parameterization is only valid in the range $0.04 < \beta < 1$. Outside of this range, the code would give parameterizations flat in $\beta$. For illustrative and practical purposes, we fitted the quark and gluon parameterization to a simple polynomial form with a fair accuracy:
\begin{eqnarray}
&&\beta f_{q/\Pomeron}(\beta,Q_{0}^2)=0.003976-0.02708 \beta + 0.3284 \beta^2 -0.7832 \beta^3+ 0.8080 \beta^4-0.3267 \beta^5 \ , \nonumber\\
&&\beta f_{g/\Pomeron}(\beta,Q_{0}^2)=0.6105-1.709 \beta + 9.873 \beta^2 -26.49 \beta^3+ 31.71 \beta^4-13.87 \beta^5 \ .
\end{eqnarray}
The Pomeron flux in this model was taken in the following form
\begin{equation}
f_{\Pomeron}(x_{\Pomeron})=-1.92308\Big(x_{\Pomeron}^{0.886}-0.0100518x_{\Pomeron}^{1.406}\Big)\Big /\Big(x_{\Pomeron}^{2.292}(-8.84615+\ln(x_{\Pomeron}))\Big) \  .
\end{equation}

\section{Parameterization of the nuclear one-body density}
\label{appendix:b}

\TABULAR[ht]{|c|c|c|}{
\hline
 & $\rho_{0}$ (fm$^{-3})$ & $c$ (fm) \\
\hline
 $^{12}$C & 0.013280 & 2.2486\\
 $^{32}$S & 0.0049717 & 3.3663\\
 $^{40}$Ca & 0.0039769 & 3.6663\\
 $^{110}$Pd & 0.0014458 & 5.308\\
 $^{197}$Au & 0.0008073 & 6.5157\\
 $^{208}$Pb & 0.0007720 & 6.6178\\
\hline}{
The parameters entering the nuclear one-body density, $\rho_A (\vec{b},z)$, of Eq.~(\ref{fermi}).}

The nuclear one-body  density $\rho_{A}(\vec{b},z)$, which appears in Eqs.~(2.2), (2.4), (2.6), (2.14)-(2.16), (3.3), (4.1) and (4.7)
 was parameterized in a two-parameter Fermi form
\begin{equation}
\rho_{A}(\vec{b},z)=\frac{\rho_{0}}{1+\exp\big[(r-c)/a\big]} \ , 
\label{fermi}
\end{equation}
where $r=\sqrt{|\vec{b}|^2 +z^2}$ and
$a=0.545$ fm and the parameters $\rho_{0}$ and $c$ are presented in Table~2. Also note that $\rho_{A}(\vec{b},z)$ is normalized as $2 \pi \int^{\infty}_{0} d|\vec{b}| \int^{\infty}_{-\infty} dz |\vec{b}| \rho_{A}(\vec{b},z)=1$.

\bigskip
\bigskip
\bigskip

\end{appendix}

\end{document}